\documentclass{elsart}
\usepackage{natbib,graphicx,times,amsmath}

\begin{document}

\newcommand{\reference}{{\sc Reference}}
\newcommand{\example}{our example problem}
\newcommand{\model}{the atmospheric model}
\newcommand{\RT}{Rossby traveling wave}
\newcommand{\RTs}{Rossby traveling waves}
\newcommand{\pair}{pair}
\newcommand{\pairs}{pairs}
\newcommand{\bea}{\begin{eqnarray}}
\newcommand{\eea}{\end{eqnarray}}
\newcommand{\bse}{\begin{subequations}}
\newcommand{\ese}{\end{subequations}}
\renewcommand{\sec}[1]{Section \ref{#1}}
\newcommand{\appndx}[1]{Appendix \ref{#1}}
\newcommand{\eq}[1]{(\ref{#1})}
\newcommand{\fg}[1]{Figure \ref{#1}}
\newcommand{\tbl}[1]{Table \ref{#1}}
\newcommand{\fginsrt}[1]{\marginpar{[Fig.\ref{#1}]}}
\newcommand{\tblinsrt}[1]{\marginpar{[Tab.\ref{#1}]}}
\newcommand{\bbR}{{\rm I\!R}}
\newcommand{\ddrv}[2]{\frac{d {#1}}{d {#2}}}
\newcommand{\pdrv}[2]{\frac{\partial {#1}}{\partial {#2}}}
\newcommand{\vctr}[1]{{\bf {#1}}}
\newcommand{\mtrx}[1]{{\bf {#1}}}
\newcommand{\vctrg}[1]{{\mbox{\boldmath${#1}$\unboldmath}}}
\newcommand{\rf}[1]{{\bar {#1}}}
\newcommand{\dgr}{$^{\circ}$}
\newcommand{\spa} {^{{\rm (a)}}}
\newcommand{\sps} {^{{\rm (u)}}}
\newcommand{\spE} {^{E}}
\newcommand{\spC} {^{C}}
\newcommand{\spu} {^{(\vctr{u})}}
\newcommand{\spp} {^{(\phi)}}
\renewcommand{\spE}{}
\newcommand{\spstr} {^{*}}
\newcommand{\spdgr} {^{\dagger}}
\newcommand{\sbmd}[1]{_{{#1}}}
\newcommand{\sbwv}[1]{_{({#1})}}
\newcommand{\sbz} {_{0}}
\newcommand{\sbzb} {_{(0)}}
\newcommand{\sbn} {_{n}}
\newcommand{\sbnb} {_{(n)}}
\newcommand{\imax} {i_{{\rm max}}}
\newcommand{\bcdt} {\diamond}
\newcommand{\sbd}  {_{\bcdt}}
\newcommand{\sbi} {_{i}}
\newcommand{\sbidp} {_{i.p}}
\newcommand{\sbipdp} {_{i+1.p}}
\newcommand{\sbiipdp} {_{[i,i+1].p}}
\newcommand{\sbip} {_{i+1}}
\newcommand{\sbiip} {_{[i,i+1]}}
\newcommand{\sbtn} {_{2n}}
\newcommand{\sbtnn} {_{2n-1}}
\newcommand{\sbijp} {_{[i,i+1],j:j+1}}
\newcommand{\vx}  {\vctr{x}}
\newcommand{\vu}  {\vctr{u}}
\newcommand{\psim}{\rf{\psi}}
\newcommand{\um}{\rf{u}}
\newcommand{\vm}{\rf{v}}
\newcommand{\qm}{\rf{q}}
\renewcommand{\um}{\rf{u}}
\renewcommand{\vm}{\rf{v}}
\newcommand{\vum}{\rf{\vu}}
\newcommand{\vxm}{\rf{\vx}}
\newcommand{\vua} {\tilde{\vctr{u}}}
\newcommand{\chia}{\tilde{\chi}}
\newcommand{\phia}{\tilde{\phi}}
\newcommand{\psia}{\tilde{\psi}}
\newcommand{\Lu}{\rf{L}\spu}
\renewcommand{\Lu}{\rf{L}}
\newcommand{\Tu}{\rf{T}\spu}
\newcommand{\Lui}{L\spu\sbi}
\newcommand{\Tui}{T\spu\sbi}
\newcommand{\Lpi}{L\spp\sbi}
\newcommand{\Tpi}{T\spp\sbi}
\newcommand{\sCidp}{s\spC\sbidp}
\newcommand{\SCi}{S\spC\sbi}
\newcommand{\SCidp}{S\spC\sbidp}
\newcommand{\GCi}{\Gamma\spC\sbi}
\newcommand{\GCidp}{\Gamma\spC\sbidp}
\newcommand{\gCi}{\gamma\spC\sbi}
\newcommand{\gCidp}{\gamma\spC\sbidp}
\newcommand{\C}{C}
\newcommand{\Cs}{C\sps}
\newcommand{\Ca}{C\spa}
\newcommand{\vxCm}{{\bar \vx}\spC}
\newcommand{\lC}{l\spC}
\newcommand{\Tma}{\rf{T}\spa}
\newcommand{\mC} {m\spC}
\newcommand{\aC} {a\spC}
\renewcommand{\aa} {a\spa}
\newcommand{\as} {a\sps}
\newcommand{\rC} {r\spC}
\newcommand{\muC} {\mu\spC}
\newcommand{\mua} {\tilde{\mu}}
\newcommand{\muCa} {\tilde{\mu}\spC}
\newcommand{\muaa} {\tilde{\mu}\spa}
\newcommand{\musa} {\tilde{\mu}\sps}
\newcommand{\azb}{a\sbzb}
\newcommand{\enb}{\epsilon\sbnb}
\newcommand{\knb}{k\sbnb}
\newcommand{\lnb}{l\sbnb}
\newcommand{\bnb}{b\sbnb}

\begin{frontmatter}



\title{The Role of Variability in Transport for Large-Scale Flow Dynamics}
\author[UMD]{Kayo Ide}\ead{ide@umd.edu}
\ead[url]{http://www.atmos.umd.edu/$\sim$ide}
 and
\author[UOB]{Stephen Wiggins} \ead{S.Wiggins@bris.ac.uk}
\ead[url]{http://www.maths.bris.ac.uk/people/faculty/maxsw/}
\address[UMD]{Department of Atmospheric and Oceanic Science, \\
Center for Scientific Computation and Mathematical Modeling, \\
Earth System Science Interdisciplinary Center, \\ \&
Institute for Physical Science and Technology, \\ 
University of Maryland, College Park, USA}
\address[UOB]{School of Mathematics, University of Bristol, Bristol BS8 1TW, UK}

\maketitle
\begin{abstract}
We develop a framework to study the role of variability in transport
across a streamline of a reference flow. 
Two complementary schemes are presented:
a graphical approach for individual cases, and
an analytical approach for general properties.
The spatially nonlinear interaction of dynamic variability 
and the reference flow results in flux variability.
The characteristic time-scale of the dynamic variability and the
length-scale of  the flux variability in a unit of flight-time
govern the spatio-temporal interaction that leads to transport.
The non-dimensional ratio of the two characteristic scales
is shown to be a a critical parameter.
The pseudo-lobe sequence along the reference streamline
describes spatial coherency and temporal evolution of transport.
For finite-time transport from an initial time up to the present,
the characteristic length-scale of the flux variability regulates
the width of the pseudo-lobes.
The phase speed of pseudo-lobe propagation averages the reference
flow and the flux variability. 
In contrast, for definite transport over a fixed time interval
and spatial segment,  the characteristic time-scale of the dynamic
variability regulates the width of the
pseudo-lobes.
Generation of the pseudo-lobe sequence appears to be synchronous with the dynamic
variability, although it propagates with the reference flow.
In either case, the critical characteristic ratio is found
to be one, corresponding to a resonance of the flux variability
with the reference flow.
Using a kinematic model,
we demonstrate the framework for two types of transport in a blocked flow 
of the mid-latitude atmosphere: across the meandering jet axis 
and between the jet and recirculating cell.
\end{abstract}

\end{frontmatter}

\newpage

\tableofcontents

\newpage

\section{Introduction}
\label{sec:intro}

\subsection{Geophysical flows and variability}
\label{sec:intro_geo}

Large-scale planetary flows are approximately
two-dimensional. 
Quite often their time evolution may be described as unsteady
fluctuations around a prominent time-averaged  structure.
Instantaneous flow fields for the velocity $\vu$ and a flow property $q$ at
time $t$ in  two-dimensional $\vx=(x,y)$ space can be written as:
\bse\label{eq:dyn}
\bea 
\vu(\vx,t)&=&\vum(\vx)+\vu'(\vx,t)~,\label{eq:dyn_u}\\
q(\vx,t)&=&\rf{q}(\vx)+q'(\vx,t)~ \label{eq:dyn_q}
\eea
\ese
where $\rf{\{\cdot\}}$ and  ${\{\cdot\}}'$ stand for time-averaged
(``reference'') and residual fluctuation (``anomaly'' or
``transient eddy'') fields,
respectively. 
The flow property $q$ here collectively represents possible passive tracers;
examples are (potential) temperature, (potential) vorticity,
chemical concentrations, humidity in the atmosphere and salinity in
the ocean. 
Usually, the flow dynamics is given by a time series of the
instantaneous fields. 
In contrast, transport is a time-integrated phenomenon.
The main goal of this paper is to identify general properties of
transport by connecting the flow dynamics and transport systematically.
To achieve this, we take  hierarchical steps which combine
a spatio-temporal analysis for the anomaly field
with a geometric method for quantifying  transport.

Transport depends significantly on the flow geometry.
Hence, we start by describing the basic nature of the flow geometry used
in this study. 
Given $\vu(\vx,t)$, streamlines to which $\vu(\vx,t)$ is locally
tangent describe the instantaneous flow geometry.
If the flow is incompressible, then the streamlines are the contours of
a streamfunction $\psi(\vx,t)$ that satisfies
$\vu(\vx,t)=[-\pdrv{}{y},\pdrv{}{x}]\psi(\vx,t)$. 
However, we make no assumption  concerning incompressibility,
but we still use a streamline field $\chi(\vx,t)$ to describe the flow
geometry:

\bea \label{eq:dyn_cji}
\vu(\vx,t)\wedge \left[\pdrv{}{x},\pdrv{}{y}\right]\chi(\vx,t)&=&0~.
\eea
By this general definition, a streamline coincides with 
a contour of $\chi(\vx,t)$.
Because we do not use the functional form of $\chi(\vx,t)$ 
to derive any mathematical formulae, we impose no specification 
on $\chi(\vx,t)$
other than the direction of the vector 
$[-\pdrv{}{y},\pdrv{}{x}]\chi(\vx,t)$ to be consistent with
$\vu(\vx,t)$.
If the flow is incompressible, 
then $\psi(\vx,t)$ can be used as $\chi(\vx,t)$.
Remarks concerning incompressibility are provided throughout the paper
using $\psi(\vx,t)$, since large-scale planetary flows can 
often be treated as incompressible.

More often than not, anomaly fields of large-scale planetary flows 
exhibit significant spatio-temporal coherency called ``variability.''
An anomaly velocity field is typically represented  as a finite linear sum of
$\imax$  modes and noise, i.e.,
$\vu'(\vx,t)=\sum_{i=1}^{\imax}\vua\sbi(\vx,t)+\mbox{ noise}$.
The $i$-th ``dynamic mode'' may be written as a spatio-temporal
decomposition:

\bea\label{eq:anomaly}\label{eq:anomaly_u} 
\vua\sbi(\vx,t)&=&\sigma\sbi~\vu\sbi(\vx)~f\sbi(t)~,\label{eq:anomaly_um} 
\eea
where $\tilde{\{\cdot\}}$ stands for spatial-temporal decomposition from
here on.
A  commonly used technique for such a spatio-temporal decomposition
is an empirical orthogonal function, or principal component
(PC) analysis, based on the covariance matrix of the anomaly field. 
Another spatio-temporal decomposition technique uses spectral
analysis, such as a normal mode analysis decomposition \citep{EIK92}.
In general,  the spatial PC $\vu\sbi(\vx)$ is normalized
over the entire flow domain so that it averages to zero and the 
norm 
is the same for any $i$.
Similarly, the temporal PC $f\sbi(t)$ is normalized over the entire time
interval.
Hence, the (ordered, positive) variance $\sigma\sbi$ with 
$\sigma\sbi\geq\sigma\sbip>0$ reflects the statistical significance of
mode $i$. 

We introduce here some basic properties of spatio-temporal coherency
on which we develop the framework to study the role of variability in transport.
As a single dynamic mode, $\chia\sbi(\vx,t)$ describes  a standing
geometry in $\vx$ which pulsates in $t$ with $f\sbi(t)$, where
$\chi\sbi(\vx)$ is the streamline field of $\vu\sbi(\vx)$.
Quite often $\chi\sbi(\vx)$ consists of coherent structures
which we call ``dynamic eddies.''
A positive eddy corresponds to a locally anti-cyclonic flow around 
a maxima  of $\chi\sbi(\vx)$.
A negative eddy corresponds to a cyclonic flow.
We define the dynamic characteristic length-scale
$\Lui$ by the typical width of the dynamic eddies in $\chi\sbi(\vx)$.
Temporal coherency is described by the
positive and negative phases of $f\sbi(t)$ based on the sign.
In this study, we define a characteristic time-scale $\Tui$
by individual intervals of the phases:
one recurrent cycle takes $2\Tui$.
If $f\sbi(t)$ is regular, then $\Tui$ is constant
and the phase condition can be given by:

\bea
f\sbi(t)\approx-f\sbi(t+\Tui)~. \label{eq:ft}
\eea
If irregular, then $\Tui$ may be a function of $t$
and $f\sbi(t)$ may possibly be described by a linear sum of
several regular components.
For simplicity, we proceed with a regular  $f\sbi(t)$ assumption. (In this paper, by the term ``regular'', we mean periodic.)
Additional comments on irregular $f\sbi(t)$ are provided in later
sections.

In the large-scale atmospheric and oceanic flows,
dominant modes with significant variance tend to have larger characteristic
scales (i.e., $\Lui\geq L\spu\sbip$ and $\Tui\geq
T\spu\sbip$ for $\sigma\sbi\geq\sigma\sbip$).
Variability  may be described by the recurrent time interval 
$2\Tui$ and geographic location of dynamic eddies.
Because of its role in the understanding of the 
atmospheric general circulation and in extended-range weather
forecasting,
low-frequency variability of the eastward jet in the mid-latitude
atmosphere has attracted significant interest over  several
decades \cite[and references therein]{Tetal01}.
Using a kinematic model, we study the role of the
variability in transport for a blocked atmospheric flow
as a demonstration of our methodology.

If a mode pair $\vua\sbi(\vx,t)$ and $\vua\sbip(\vx,t)$ satisfies certain
conditions,  then dynamic eddies  in the corresponding streamfunction field

\bea\label{eq:uiip}
\chia\sbiip(\vx,t)&\equiv&\chia\sbi(\vx,t)+\chia\sbip(\vx,t)~:
\eea
can exhibit recurrent evolution as follows.
A \pair\  has approximately equal variance and
characteristic scales.
Spatially, there are some sets of dynamic eddies in  
$\chi\sbi(\vx)$ and $\chi\sbip(\vx)$ which align along  
common curves in $\vx$, with their centers staggered with respect to each other.
Temporally,  
individual phases of :$f\sbi(t)$ and $f\sbip(t)$ are lag-correlated

\bea
f\sbi(t)\approx f\sbip(t+\frac{\Tui}{2})~, \label{eq:fiip} 
\eea
whether they are regular or not.
We always choose the first mode $i$ so that
$f\sbi(t)$ precedes $f\sbip(t)$.
Hence, dynamic eddies in $\chia\sbiip(\vx,t)$
evolve and recur along these alignment curves.
The phase speed $b\spu\sbiip$ of the dynamic eddies is $\Lui/\Tui$
because they move a distance $2\Lui$ over one recurrent cycle time
$2\Tui$. 
The four phases of $\chia\sbiip(\vx,t)$ at every ${\Tui}/{2}$ are
$\chi\sbi(\vx)$, $\chi\sbip(\vx)$, $-\chi\sbi(\vx)$,  and
$-\chi\sbip(\vx)$  in a time sequence.
We call such a mode pair the ``dynamically coherent pair $[i,i+1]$,''
or simply ``pair,''  denoted by paired subscripts in brackets. 
The spatial and temporal conditions mentioned above govern the
transport mechanism by a \pair.

A spatio-temporal decomposition of the form \eq{eq:anomaly_u}
arises also from numerical modeling of large-scale planetary 
flow dynamics using a spectral method.
Given a set of pre-selected spatial modes $\vu\sbi(\vx)$,
$\sigma\sbi f\sbi(t)$ can be given as the residual of
the spectral coefficients around the time average.
Therefore, an individual spectral mode can be treated as one dynamic
mode for studying its role in transport.
If several spectral coefficients share the same temporal 
spectra, then they can be linearly rearranged into a set of
dynamic modes which better describe the variability.

\subsection{Transport}
\label{sec:intro_transport}

Transport issues arise in a number of different settings  in the climate system.
For example, heat and water exchanges at the interface of the atmosphere
and ocean are  important for maintaining the earth's energetics.
The streamwise transport of momentum, energy and other physical
properties are important elements of the  atmospheric and oceanic
general circulation.

In this study, we focus on the coherency of spanwise transport 
across a reference streamline due to variability.
If the flow dynamics has no variability, then no transport,  except via
molecular diffusion, occurs across a reference streamline.
Unsteady fluctuations stir the flow and induce kinematic transport. 
Lagrangian lobe dynamics is a deterministic
technique which computes fluid particle transport between two
kinematically distinct regions in an unsteady flow \citep{W92}.
Another branch of transport theory uses stochastic models 
and describes material transport by the random motion of fluid
particles  
\cite[and references therein]{bm1, bm2, bm3}.
The combined effects of molecular diffusion and kinematic advection
can be studied by treating the unsteady fluctuation at the high- and
low-frequency limits  \citep{RP99}.
These methods take the Lagrangian view:
to obtain properties associated with transport,
they follow individual particles according to:

\bea \label{eq:udxdt}
\ddrv{}{t}\vx = \vu (\vx,t)~.
\eea

Flow geometry plays a critical role in transport.
Lagrangian lobe dynamics uses the geometric approach of dynamical
systems theory.
It relies on both an unstable manifold from an upstream
distinguished hyperbolic  trajectory (DHT) and a stable  
manifold from a downstream DHT in the unsteady flow.
If these manifolds intersect, then a series of Lagrangian 
lobes containing fluid particles becomes a deformable boundary.   
Lobe-by-lobe, the ``turnstile'' mechanism transports fluid particles
between the regions as the lobes advect downstream.
Over the past decade, Lagrangian transport theory has been
applied to many geophysical transport problems;
see \cite*{warfm, physrep} for a review of  lobe dynamics applications in geophysical flows.

Using the geometric approach of dynamical systems theory,
\cite*{IW_TIMEI, IW_TIMEII} recently developed a parallel formulation
of the Transport Induced by Mean-Eddy Interaction (TIME) theory.
Unlike Lagrangian lobe dynamics,
TIME can compute transport of 
fluid particles and flow properties across any boundary 
defined by a reference streamline.   
When applied to a separatrix connecting upstream and downstream DHTs,
The TIME
gives a leading order approximation to Lagrangian lobe dynamics. 

A critical distinction between these two transport theories is that 
TIME uses the  interaction of the anomaly velocity with the reference flow 
as in \eq{eq:dyn_u}, while the Lagrangian uses the full
velocity as in \eq{eq:udxdt}. 
Therefore,  TIME is  natural for studying
the role of variability in transport.
Using an idealized kinematic model of a large-scale atmospheric flow
associated with Rossby traveling waves, we demonstrate our method
throughout this paper.
The kinematic model has a reference flow similar to  that in the
\cite{CdV79} model
based on the  dynamic quasi-geostrophic equations with topography.
The dynamic model has frequently been used to study low-frequency
variability of atmospheric dynamics 
\cite[and references therein]{Tetal01}. 
The idealized model has been used to study Lagrangian transport by 
\cite{P91} for
chaotic mixing of particles and tracers, and by \cite{MW98} using
Lagrangian lobe dynamics.
This study emphasizes the role of variability and
deepens our understanding of the transport mechanism 
as a spatio-temporal interaction of the reference meandering jet
and the Rossby traveling waves.

This paper is organized as follows.
In \sec{sec:model}, we briefly describe the model.
In \sec{sec:et}, we schematically 
present the basic ideas and formulae of the TIME.
In \sec{sec:flx2et}, we connect dynamic variability 
to transport step by step.
In \sec{sec:graphic}, we use a graphical approach for
individual cases.
In \sec{sec:analytic} and \appndx{appndx:analytic},
we explore the general properties of the TIME
and the role played by the variability using the analytic
approach.
Finally we give concluding remarks in \sec{sec:cncl}.

\section{Atmospheric model}
\label{sec:model}

The model flow is incompressible in the $x$-periodic channel domain.
The reference streamfunction:
\bea
\psim(\vx)&=&\azb\sin(\pi x)\sin(\pi y) - y 
 \label{eq:rossby_psim} 
\eea
satisfies the steady state condition of the dynamic quasi-geostrophic
(QG) model for large-scale planetary flows, 
$\pdrv{}{t}\qm(\vx,t)=0$,
where $\qm(\vx)=\nabla_{\vx}^{2}\psim(\vx)+y$ is the potential 
vorticity of the reference flow.
The first term of the right-hand side is
the principal Rossby wave with 
amplitude $\azb$ and wave-number vector $(1,1)$ in $\vx$,
which consists of the upstream anti-cyclonic and downstream  cyclonic
recirculating cells (\fg{fg:rt_rf}a).
The second term is a uniform jet induced by 
moving with the principal Rossby wave  relative to the earth-fixed
frame at a constant phase speed $1$  (\fg{fg:rt_rf}b).
\fginsrt{fg:rt_rf}
The geometry of the reference flow \eq{eq:rossby_psim} bifurcates
as the amplitude $\azb$ varies.
In this study, we consider a case where the flow is supercritical,
i.e., $1>\azb>{1}/{\pi}$, so that it corresponds to a blocked state of
the atmospheric jet over  topography \citep{CdV79}.
If the flow field is given by a numerical simulation of the 
QG model, then the time average of $\psi(\vx,t)$ can be
used as $\psim(\vx)$.
In a blocked flow, the two separatrices divide the reference flow field 
into three kinematically distinct regions:
a pair of upstream and downstream recirculation cells and an
eastward jet (\fg{fg:rt_rf}c).
The jet flows faster where it makes turns at the trough $(x=1/2)$ and
ridge $(x=3/2)$ to pass the recirculating cells.

In this kinematic model,
$N$ additional \RTs\  comprise the anomaly field
$\psi'(\vx,t)=\sum_{n=1}^{N} \psi\sbnb(\vx,t)$,
where the $n$-th wave
\bea\label{eq:rossby_psia_i}
\psi\sbnb(\vx,t)&=&\enb\sin[\knb\pi (x-\bnb t)]\sin(\lnb\pi y)~:
\eea
is defined by  amplitude $\enb$ 
and wave-number vector $(\knb,\lnb)$  for $\knb\lnb>1$.
Because a Rossby wave with a higher wave number 
travels faster, the phase speed 
$\bnb\equiv 1-{2}/({\knb^{2}+\lnb^{2}})$ is positive.
Throughout the paper, the subscript in parentheses represents
the identity of the \RT.
\tbl{tbl:rt} summarizes the five additional \RTs\
used in this study.
\tblinsrt{tbl:rt}
For  comparison purposes,
we fix the amplitude $\enb=0.1$ for all $n$.

Based on the spatio-temporal decomposition
(\sec{sec:intro_geo}), 
the $n$-th  \RT\ is made up by a \pair\ $[2n-1,2n]$,
i.e., $\psi\sbnb(\vx,t)\equiv\psia_{[2n-1,2n]}(\vx,t)$, where
the notation convention for the \pair\ follows \eq{eq:uiip}.
The variance, spatial PC, and temporal PC of
$\psia\sbtnn(\vx,t)$ and $\psia\sbtn(\vx,t)$ corresponding to
\eq{eq:anomaly} can be defined as:
\bea\label{eq:rt_mode}
\begin{array}{lllclrr}
(\sigma\sbtnn,&\psi\sbtnn(\vx),&f\sbtnn(t))
&=& (\enb,&\sin(\knb\pi x)\sin(\lnb\pi y),&\cos(\knb\bnb\pi t)~)~, \\
(\sigma\sbtn,&\psi\sbtn(\vx),&f\sbtn(t))
&=& (\enb,&-\cos(\knb\pi x)\sin(\lnb\pi y),&\sin(\knb\bnb\pi t)~)~.
\end{array}
\eea
In the spectral representation of the dynamic QG model,
$\sigma\sbi f\sbi(t)$ can be thought of as the residual of the 
spectral coefficients for pre-selected $psi\sbi(\vx)$.

Variability of this model has the properties commonly
observed in large-scale planetary flows as discussed in
\sec{sec:intro_geo}, if we choose 
$\enb\geq\epsilon\sbwv{n+1}$, $\knb\leq k\sbwv{n+1}$, 
and $\lnb\leq l\sbwv{n+1}$ with $\knb\lnb<k\sbwv{n+1}l\sbwv{n+1}$.
The spatial PCs of \pair\ $[2n-1,2n]$ have  $\lnb$ latitudinal lines
along which $2\knb$  dynamic eddies align (\fg{fg:rt_psia}).
\fginsrt{fg:rt_psia}
The characteristic length-scale is $L\spu\sbtnn={1}/{\knb}$ 
along the dynamic alignment curve.
The temporal PCs 
have the characteristic time-scale  $T\spu\sbtnn=1/(\knb\bnb)$.
Pair by pair, $L\spu\sbtnn$ and $T\spu\sbtnn$ decreases
as $\knb\lnb$ increases.

Because the coherent evolution of dynamic eddies is essential in understanding
the role of variability in transport,
we briefly demonstrate how a \pair\ generates a \RT\
by following \sec{sec:intro_geo}.
The spatial PC of the first dynamic mode $2n-1$ has $\lnb$ sets of 
$2N\spu\sbtnn=2\knb$
dynamic eddies (left panels of \fg{fg:rt_psia})
which alternate in sign along the dynamic alignment curves
$y=(j-\frac{1}{2})/\lnb$ for $j=1,\ldots,\lnb$.
The spatial PC of the second dynamic mode  $2n$ also has
$\lnb$ sets of $2N\spu\sbtnn$ dynamic eddies along the same curves
(middle panels of \fg{fg:rt_psia}).
The two spatial PCs have the centers of the dynamic eddies 
staggered with respect to each other along the alignment curves,
while their corresponding temporal PCs satisfy \eq{eq:ft} and
\eq{eq:fiip}.
The four phases of  $\psia\sbmd{[2n-1,2n]}(\vx,t)$
over one recurrent cycle $2T\spu\sbtnn$ 
are $\psi\sbtnn(\vx)$, $\psi\sbtn(\vx)$, $-\psi\sbtnn(\vx)$,
and $-\psi\sbtn(\vx)$ in a time sequence  (\fg{fg:rt_psia}).
In summary, $\psia_{[2n-1,2n]}(\vx,t)$ has $\lnb$ sets of  $N\spu\sbtnn$
positive and $N\spu\sbtnn$ negative dynamic eddies that 
propagate straight eastward with the positive phase speed 
$b\spu\sbmd{[2n-1,2n]}=L\spu\sbtnn/T\spu\sbtnn=\bnb$.

It is worth noting the case where the spatial PCs
of a \pair\ have an opposite phase relation, i.e.,
$\psi\sbtnn(x,y)=\psi\sbtn(x-{L\spu\sbtnn}/{2},y)$
instead of $\psi\sbtnn(x,y)=\psi\sbtn(x+{L\spu\sbtnn}/{2},y)$
as in \eq{eq:rt_mode}.
In $\psia_{[2n-1,2n]}(\vx,t)$,
$\lnb$ sets of $2N\spu\sbtnn$ dynamic eddies 
would advect in the reverse direction with a negative constant phase
speed $-b\spu\sbmd{[2n-1,2n]}$  along the same curves.
This artificial setting could happen if 
a \RT\ with smaller length-scale travels slower in the
earth-fixed frame.

\section{Basics of TIME}
\label{sec:et}

We briefly present basic ideas and formulae of  the TIME used in later sections.
The mathematical derivations and a detailed discussion can be found in 
\cite*{IW_TIMEI, IW_TIMEII}.

\subsection{Boundary of transport}
\label{sec:et_C}
In TIME, we evaluate transport across
a stationary boundary defined by a reference
streamline $\rf{\chi}(\vx)$ of our choice. 
We denote this boundary curve by $C$.
To present the theory using the geometric approach of dynamical systems,
we define two  coordinate variables along $C=\{\vxCm(s)\}$.
Here $s$ is the flight-time coordinate in  a unit of time
so that a fluid particle starting from 
$\vxCm(s\spstr-\triangle t)$ reaches $\vxCm(s\spstr)$
after a time interval $\triangle t$ in the reference flow
(\fg{fg:string}a).
\fginsrt{fg:string}
By this definition, $s$ satisfies $\ddrv{}{s}\vxCm(s)=\vum(\vxCm(s))$
and any particle travels with a non-dimensional 
phase speed $\ddrv{}{t}s=1$.

We denote the upstream  and downstream end points of $C$ by $s_{-}$
and $s_{+}$, respectively.
The length  of $C$ measured in a unit of (flight) time
depends on the kinematic type of $C$ as follows.
If $C$ is an unstable or stable invariant manifold associated with
a DHT (i.e., a hyperbolic stagnation point in the reference flow), then
it has a semi-infinite length with $s_{-}\to-\infty$ or
$s_{+}\to\infty$ towards the direction of the DHT.
If $C$ is a separatrix along which an unstable and stable invariant
manifold coincide, then it has a bi-infinite length.
If $C$ has no DHT at either end point,
then it has a finite length.
Using an example (\fg{fg:rt_rf}), this study demonstrates our method 
across two types of $C$:
an infinite length along the upstream separatrix $\Cs$ 
between the jet and the recirculating cell, and
a finite length along the jet axis $\Ca$ between the northern and
southern parts of the jet.

Arc-length $l=\lC(s)$ is another coordinate variable along $C$
corresponding to the arc-length distance traveled by a particle.
The geometry of TIME can be naturally described using $l$
because it has a unit of length.
The two coordinate variables are related by the local velocity, i.e.,
$\ddrv{}{s}\lC(s)=|\vum(\vxCm(s))|$.
From here on, we let $(s,t)$ and $(\lC(s),t)$ represent the
combination of a position $\vxCm(s)$ along  $C$ at a time $t$.

\subsection{Mechanism of TIME}
\label{sec:et_mech}
\begin{figure}[htb!]
\begin{center}
\includegraphics[height=6.cm]{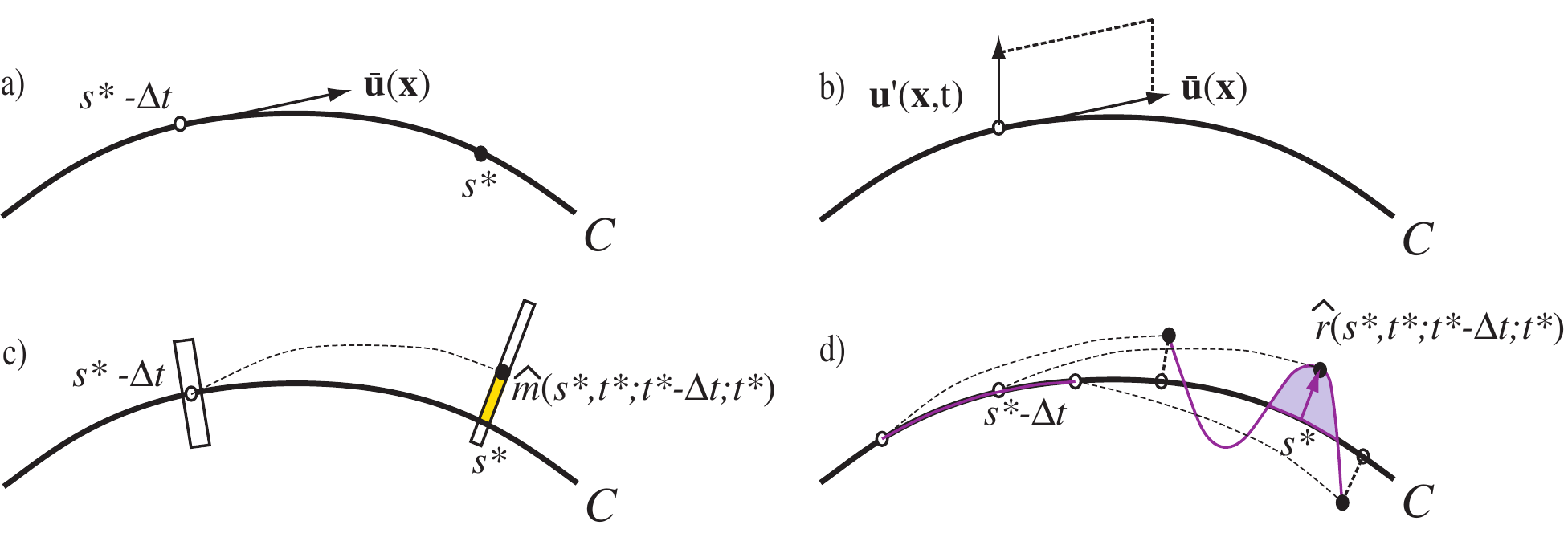}
\end{center}
\caption{Geometry associated with the TIME:
a) boundary curve $C$ and particle motion in the steady reference
   flow;
b) instantaneous flux across $C$ in the unsteady flow;
c) fluid column evolution in the unsteady flow and accumulation of
   flow property;
d) geometry of particle transport $C$ and pseudo-lobe defined by
   $R$ and $C$.}
\label{fg:string}
\end{figure}

In an unsteady flow, a component of instantaneous velocity normal to
$C$ repels particles away from $C$.
The instantaneous flux normal to $C$ at $(s,t)$

\bea\label{eq:mu} 
\muCa(s,t) &\equiv&\vum(\vxCm(s))\wedge\vu'(\vxCm(s),t)~:
\eea
is the signed area of the parallelogram defined by $\vum(\vxCm(s))$
and $\vu'(\vxCm(s),t)$  (Figure \ref{fg:string}b),
where ``$\wedge$'' denotes the normal wedge product.
For  $\muCa(s,t)>0$,  the instantaneous flux is from right to left 
across $C$ with respect to the forward direction of the reference
flow.
The direction of the flux reverses for  $\muCa(s,t)<0$.

There are two approaches to the TIME.
One considers the net transport of the flow property $q$ defined in
\eq{eq:dyn_q}. 
This is pursued by accumulating instantaneous flux while 
advecting with the flow.
Another approach determines the geometry associated with particle transport.
This is pursued by computing the displacement distance of a particle
originally placed on $C$.
In the special case of mass transport with $q=1$ in an
incompressible flow, the two approaches lead to the same formula.

We illustrate the first approach 
using a fluid column that originally
intersects with $C$ at $(s\spstr-\triangle t, t\spstr-\triangle t)$
(marked by a white circle in Figure \ref{fg:string}c).
If the flow is steady, then that original intersection 
remains on $C$ at $(s\spstr-t\spstr+\tau,\tau)$ 
for $t\spstr-\triangle t<\tau<t\spstr$ and 
reaches $(s\spstr,t\spstr)$ after a time interval $\triangle t$.
In the unsteady flow,
the original intersection moves away from $C$ due to the
component of $\vu'(\vx,t)$ normal to $C$ (marked by a dark circle).
The displaced portion of the column   (marked by a shaded area) stores
continuously accumulating instantaneous flux of $q$ across $C$.
This accumulation of $q$ is the TIME we wish to compute.
It can be shown that the net accumulation of $q$  in the column over the
time interval is:  
\bea 
\hat{m}(s\spstr,t\spstr;t\spstr-\triangle t:t\spstr)
 =\int_{t\spstr-\triangle t}^{t\spstr}
\rf{q}(\vxCm(s\spstr-t\spstr+\tau)) \muCa(s\spstr-t+\spstr\tau,\tau) 
 d\tau~, \label{eq:mp}
\eea
up to the leading order, where the integrand
is instantaneous flux of $q$ at $(s\spstr-t\spstr+\tau,\tau)$
for $\tau\in[t\spstr-\triangle t,t\spstr]$.

In the illustration above, TIME is evaluated over the time
interval just past the evaluation time $t\spstr$.
A straightforward extension of  \eq{eq:mp} gives
the TIME of $q$ evaluated at an arbitrary $(s,t)$ that occurs 
over a spatial segment $[s_{a}, b_{b}]$ 
during a time interval  $[t_{0}, t_{1}]$:
\bea
\mC(s,t;s_{a}:s_{b},t_{0}:t_{1})
 =\int_{t_{0}}^{t_{1}} H(s-t+\tau;s_{a}:s_{b})
  \tilde{\mu}^{C,q}(s-t+\tau,\tau)
  d\tau~, \label{eq:m}
\eea
where $\tilde{\mu}^{C,q}(s,t)\equiv\rf{q}(\vxCm(s))\muC(s,t)$.
The two limits of the integral specify the temporal interval and
\bea\label{eq:H}
H(\theta;s_{a}:s_{b})=\left\{\begin{array}{ll}
1 & \mbox{ for $s_{a}\leq\theta\leq s_{b}$}, \\
0 & \mbox{ otherwise}~: \end{array} \right.
\eea
takes care of the spatial segment.
The sign of $\mC(s,t;s_{a}:s_{b},t_{0}:t_{1})$ shows the direction of
TIME:  a positive sign corresponds to accumulation
of $q$ from the right to left region across $C$ (with respect to the forward direction of the reference flow).
The directions reverse for the negative sign.
We call $\mC(s,t;s_{a}:s_{b},t_{0}:t_{1})$ the ``accumulation
function,'' where the two sets of arguments 
$(s,t)$ and $(s_{a}:s_{b},t_{0}:t_{1})$ are
the ``evaluation point and time'' and ``domain of integration,''
respectively.

We illustrate the second approach for the geometry associated with
particle transport
using a material line $R$ initially placed along $C$  at time
$t\spstr-\triangle t$  (Figure \ref{fg:string}d).
In the unsteady flow, the normal component of the unsteady velocity
displaces $R$ from $C$.
The deformed $R$ and stationary $C$ form structures called 
``pseudo-lobes'' which  show the local coherency of TIME. 
The prefatory word ``pseudo'' is used to distinguish
from its counterpart in Lagrangian lobe dynamics. 
In Figure \ref{fg:string}d, the shaded area surrounded by 
$R$ and $C$ is a pseudo-lobe of TIME over 
the immediate past time interval $\triangle t$.
The arrow at $(s\spstr,t\spstr)$
is the corresponding signed distance 
$\hat{r}(s\spstr,t\spstr;t\spstr-\triangle t:t\spstr)$. 
In general, the ``displacement distance'' function for the
signed normal distance from $C$ to $R$ can be written as:
\bse\label{eq:da}
\bea\label{eq:da_d}
r\spC(\lC(s),t;s_{a}:s_{b}, t_{0}:t_{1})
 &=&{\aC(s,t;s_{a}:s_{b},t_{0}:t_{1})}/{||\vum(\vxCm(s))||}~,
\eea
where 
\bea
\aC(s,t;s_{a}:s_{b}, t_{0}:t_{1})&=& 
e(s:s_{a})\int_{t_{0}}^{t_{1}}
H(s-t+\tau;s_{a}:s_{b})
\tilde{\mu}^{C,e}(s-t+\tau,\tau)
d\tau~:
\label{eq:da_a}
\eea
\ese
is the so-called  ``displacement area'' function having a unit of
area per (flight) time  and taking the compressibility of the reference
flow into account:  
\bea\label{eq:da_ae}
e(s:s_{a})&=&
\exp\left\{\int_{s}^{s_{a}}J[\vum,\vx](\vxCm(\theta))d\theta\right\}~,
\eea
and $\tilde{\mu}^{C,e}(s,t)\equiv e(s_{a}:s) \muCa(s,t)$.
The convention of the arguments  follows that of 
$\mC(s,t;s_{a}:s_{b},t_{0}:t_{1})$, but $\lC(s)$ replaces $s$ for the
distance function \eq{eq:da_d} to describe the transport geometry
in a physically consistent unit of length along $C$.  
The sign of the displacement functions represents the transport
direction, as in the accumulation function.

When the displacement functions are applied to a separatrix with the
infinite domain of integration in both $s$ and $t$,
$\aC(s,t;-\infty:\infty,-\infty:\infty)$
 coincides with the
so-called ``Melnikov function'' used in Lagrangian lobe dynamics.
Accompanying 
$r\spC(\lC(s),t;-\infty:\infty,-\infty:\infty)$ 
can be interpreted as the leading order
approximation of the signed distance from the 
stable to unstable invariant manifolds measured at $(s,t)$
normal to $C$.

We call the accumulation and displacement functions in \eq{eq:m}
and \eq{eq:da} 
collectively the ``TIME functions.''
They can be evaluated  at any point and time in the $(s,t)$ space,
and represent the amount of transport that has occurred, is
occurring, or will occur, depending on the relative relation of $(s,t)$ to 
$(s_{a}:s_{b},t_{0}:t_{1})$.
We define two categories of TIME relevant to 
large-scale geophysical flows as follows.
``Definite'' TIME measures the net amount of
transport for a fixed $(s_{a}:s_{b},t_{0}:t_{1})$ which can extend to
$(-\infty:\infty,-\infty:\infty)$ if $C$ is a separatrix, as above.
In contrast, ``Finite-time'' TIME measures  
transport over the immediate past time interval starting at
time $t\spdgr$ up to the present $t$, i.e., the domain of integration 
$(s\spdgr:s,t\spdgr:t)$ varies as $t$ progresses.

\subsection{Spatial coherency and temporal evolution} 
\label{sec:et_pL}

Using,  for simplicity, the displacement area function 
$\aC(s,t)=\aC(s,t:s_{a}:s_{b},t_{0}:t_{1})$
related to particle transport,
we describe the spatial coherency of TIME based on the 
geometry of pseudo-lobes at $t$ and the temporal evolution based on 
the phase speed and deformation in $t$.
For simplicity of notation, we drop the argument for the domain of
integration.
It is straightforward to extend this discussion for TIME
of a flow property $q$ by replacing $\aC(s,t)$ with $\mC(s,t)$.
We drop the arguments corresponding to the domain of integration
for  notational convenience.

We start from spatial coherency at time $t$.
The zero sequence $\{s\spC_{j}(t)\}$ of $\aC(s,t)$ defined by
\bea\label{eq:pPIP}
\aC(s\spC_{j}(t),t)&=&0~:
\eea
is the ordered intersection sequence of $R$ with $C$ 
from upstream to downstream (\fg{fg:string}), 
because the zeros of $\aC(s,t)$ and $\rC(\lC(s),t)$ have one-to-one
correspondence from \eq{eq:da_d}.
We call $\{s\spC_{j}(t)\}$ the ``pseudo-PIP'' sequence.

The definition of  the ``pseudo-lobe'' ${\cal L}\spC_{j:j+1}(t)$ 
follows naturally in the $(s,a)$ coordinate as the region
defined by the two curves, $a=\aC(s,t)$ for $R$ and $a=0$ for $C$,
over the segment between the two adjacent pseudo-PIPs,
$s\in[s\spC_{j}(t),s\spC_{j+1}(t)]$.
The sign of $\aC(s,t)$ normally alternates pseudo-lobe by pseudo-lobe
along $C$: a positive pseudo-lobe 
with $\aC(s,t)>0$  lies on the left of $C$ and represents local
coherency of TIME from  the right across $C$.
The directions are reversed for a negative pseudo-lobe.
The definition of ${\cal L}\spC_{j:j+1}(t)$ can be extended
into the $(l,r)$ space, which can in turn be projected into the $\vx$ 
space along $C$. 
Given $\aC(s,t)$, the signed area of ${\cal L}\spC_{j:j+1}(t)$ is:
\bea\label{eq:pA}
A\spC_{j:j+1}(t) &\equiv&
\int_{\lC(s\spC_{j}(t))}^{\lC(s\spC_{j+1}(t))} r\spC(l,t)dl
=\int_{s\spC_{j}(t)}^{s\spC_{j+1}(t)} \aC(s,t)ds~.
\eea
In the $(s,a)$ space, spatial coherency of TIME
is described by the geometry of ${\cal L}\spC_{j:j+1}(t)$, i.e.,
the separation distance of pseudo-PIPs
$\triangle s\spC_{j:j+1}(t)\equiv s\spC_{j+1}(t)-s\spC_{j}(t)$
and average height
$\check{a}\spC_{j:j+1}(t)\equiv A\spC_{j:j+1}(t)/\triangle s\spC_{j:j+1}(t)$.
The latter also measures transport efficiency of  
${\cal L}\spC_{j:j+1}(t)$ in a unit of area per time.
Naturally, temporal evolution can be described
by the phase speed $\ddrv{}{t}s\spC_{j}(t)$ for propagation,
and change of $A\spC_{j:j+1}(t)$ for  deformation.
In the $(l,r)$ and $\vx$ spaces,  the geometry of 
${\cal L}\spC_{j:j+1}(t)$ can be significantly distorted  
if $\vum(\vxCm(s))\approx 0$.
The effect is most apparent along a separatrix near the DHT.

\section{Variability in instantaneous flux}
\label{sec:flx2et}

In a series of steps, we build a framework to identify the signature of dynamic
variability in transport.

\subsection{Global instantaneous flux}
\label{sec:flx2et_global}
We begin by globally examining the instantaneous flux field in  $\vx$:
\bse\label{eq:phi}
\bea \label{eq:phi_a}
\phi(\vx,t)&\equiv&\vum(\vx)\wedge\vu'(\vx,t).
\eea
Using the spatio-temporal decomposition of dynamic variability
\eq{eq:anomaly}, the  ``flux variability''  is:
\bea
\phia\sbi(\vx,t)&\equiv&\vum(\vx)\wedge\vua\sbi(\vx,t)
=\sigma\sbi\phi\sbi(\vx)f\sbi(t)~, \label{eq:phi_am}
\eea
for mode $i$ and 
$\phia\sbiip(\vx,t)\equiv\phia\sbi(\vx,t)+\phia\sbip(\vx,t)$
for \pair\ $[i,i+1]$.
The total instantaneous flux field $\phi(\vx,t)$ is a linear sum of
contributions from all modes and noise.

An expected, yet striking feature of \eq{eq:phi_am}, is that 
flux variability preserves the temporal PCs of the dynamic variability.
This is because the reference flow $\vum(\vx)$  has no temporal
component  to interact with $\vua\sbi(\vx,t)$.  
The temporal signature of dynamic variability is carried
into the flux variability as the characteristic time-scale $\Tui$.
In contrast, the spatial component of the flux mode
\bea\label{eq:phi_m}
\phi\sbi(\vx)&\equiv&\vum(\vx)\wedge\vu\sbi(\vx)~:
\eea
\ese
is the result of the nonlinear interaction between the dynamic variability
and the reference flow.
If the flow is incompressible, then
it leads to a geometric representation using the streamfunction, i.e.,
$\phi\sbi(\vx)=-J[\psim(\vx),\psi\sbi(\vx)]$.
Locally at $\vx$, $\phi\sbi(\vx)$ is the signed area of the
parallelogram defined by $\vum(\vx)$ and $\vu\sbi(\vx)$ 
(see also \fg{fg:string}c).  
Globally, it may have coherent structures which
we call ``flux eddies.''
A positive flux eddy represents local coherency of the 
instantaneous flux
from right to left across the reference streamline.
The direction of the flux reverses for a negative flux eddy.
We define the characteristic length-scale  $\Lpi$ of
$\phi\sbi(\vx)$ by the typical width of the flux eddies.

The spatial signatures of dynamic variability in $\phi\sbi(\vx)$ can be
found in the nonlinear generation of flux eddies from 
the interaction of dynamic eddies and the reference flow.
The amplitude of $\vum(\vx)$,  $\vu\sbi(\vx)$, and angle between them
define the spatially nonlinear interaction \eq{eq:phi_m}.  
Hence flux eddies tend to form where $\vum(\vx)$ is swift
and the dynamic eddies concentrate.
The center of a dynamic eddy should be near the edge of a flux eddy
because $|\vu\sbi(\vx)|=0$ and hence $\phi\sbi(\vx)=0$
there (compare \fg{fg:rt_phia_x} with
\fg{fg:rt_psia}). 
\fginsrt{fg:rt_phia_x}
Conversely, the center of a flux eddy may be located between two adjacent 
dynamic eddies because $|\vu\sbi(\vx)|$ may be large there.
As a result, flux eddies tend to stagger with respect to dynamic eddies in a
nonlinear way.
They may not distribute homogeneously
even where dynamic eddies do.
Due to the spatially nonlinear interaction, 
$L\spp\sbi$ and $L\spu\sbi$ can be different.

Using \model, we demonstrate the nonlinear generation of
flux eddies.
Because the reference flow is swift along the jet axis and
dynamic eddies distribute homogeneously everywhere, 
all significant flux eddies emerge along the reference jet only. 
The flux eddies may distribute inhomogeneously even along the jet:
strong flux eddies lie near the ridge  and 
trough where the jet flows fastest.
For the $n$-th \RT,
both $\phi\sbtnn(\vx)$ and  $\phi\sbtn(\vx)$ have
$N\spp\sbtnn$ positive and $N\spp\sbtnn$ negative flux eddies
(\tbl{tbl:rt}). 
The two strong flux eddies at the ridge and trough have the same sign
unless the wave numbers $(\knb,\lnb)$ are both even.
Because $2N\spp\sbtnn$ flux eddies emerge as the jet cuts 
across $\lnb$ horizontal lines of  $2N\spu\sbtnn$ dynamic eddies, 
Flux eddies outnumber dynamical eddies per respective
alignment curve, i.e., 
$2N\spp\sbtnn>2N\spu\sbtnn$ for $\lnb>1$.
Thus, the characteristic length-scale $L\spp\sbtnn=1/N\spp\sbtnn$ 
is smaller than  $L\spu\sbtnn=1/N\spu\sbtnn$.

Having understood the spatial features in $\phi\sbi(\vx)$
and $\phi\sbip(\vx)$,
we now describe the evolution of flux eddies in 
the spatio-temporal decomposition $\phia\sbi(\vx,t)$ and
$\phia\sbiip(\vx,t)$.
The temporal PCs $f\sbi(t)$ and $f\sbip(t)$ satisfy the 
phase conditions \eq{eq:ft} and \eq{eq:fiip}.
As a single mode, $\phia\sbi(\vx,t)$ is a geometric pattern
that pulsates with the same recurrent cycle $2\Tui$ as
$\chia\sbi(\vx,t)$.
As a pair $[i,i+1]$, $\phia\sbiip(\vx,t)$ has 
some `flux alignment curves'' along which flux eddies undergo
recurrent evolution.
The four phases of $\phia\sbiip(\vx,t)$ at every ${\Tui}/{2}$ 
are $\phi\sbi(\vx)$, $\phi\sbip(\vx)$, $-\phi\sbi(\vx)$, and
$-\phi\sbip(\vx)$ over one recurrent cycle $2\Tui$ (\fg{fg:rt_phia_x}).
The flux phase speed $b\spp\sbiip$ is $L\spp\sbi/\Tui$.
Because of the spatially nonlinear interaction, 
the evolution of flux and dynamic eddies may not exhibit a one-to-one
correspondence,  except for the same recurrent cycle:
their alignment curves and respective phase speeds can differ.

In \model,
any $\phia\sbi(\vx,t)$ has $2N\spp\sbi$ flux eddies that
pulsate with the recurrent cycle $2\Tui$.
As the $n$-th \RT\ travels straight eastward along $\lnb$
lines, it generates $N\spp\sbtnn$ positive and $N\spp\sbtnn$ negative
flux eddies that propagate eastward along the reference jet in
$\phia_{[2n-1,2n]}(\vx,t)$. 
The flux eddies  have a phase speed 
$b\spp_{[2n-1,2n]}=L\spp\sbtnn/T^{(\vu)}\sbtnn$,
which is smaller than $b\spu_{[2n-1,2n]}$ of the dynamic
eddies by a factor $N\spu\sbtnn/N\spp\sbtnn$ for $\lnb>1$.
Two flux eddies pass the ridge and trough synchronously 
when their amplitude reaches maximum.
The sign of the two is the same, unless both wave numbers $(\knb,\lnb)$ are
even. 
Strikingly, these \RTs\ inducing instantaneous flux 
near the meandering jet but not so much in the recirculation
cells, despite traveling homogeneously throughout the flow domain.

It is worth mentioning the additional case discussed in \sec{sec:model}
where the pair's spatial PCs have an opposite phase relation.
Then, the flux eddies in $\phia_{[2n-1,2n]}(\vx,t)$ would propagate in the
reverse direction  with a negative constant phase speed
$-b\spp_{[2n-1,2n]}$ along the flux alignment curve.

\subsection{Instantaneous flux across $C$}
\label{sec:flx2et_st}
The most basic element of the TIME functions 
is the instantaneous flux $\muCa(s,t)$ across $C$ as in
\eq{eq:m} and \eq{eq:da}.
Using $\phia\sbi(\vx,t)$, 
we identify the signature of dynamic variability in $\muCa\sbi(s,t)$.
Applying  \eq{eq:anomaly_u} to \eq{eq:mu} and using 

\eq{eq:phi_am} give:
\bse\label{eq:muca}
\bea\label{eq:muca_i}
\muCa\sbi(s,t)&\equiv
\sigma\sbi\muC\sbi(s)f\sbi(t)~.
\eea
The temporal component $f\sbi(t)$ directly relates to the
dynamic variability $\vua\sbi(\vx,t)$.
The spatial component 
\bea\label{eq:must_s}
\muC\sbi(s)=\phi\sbi(\vxCm(s))~:
\eea
\ese
is extracted from the global mode $\phi\sbi(\vx)$ 
in terms of flux eddies along $C$ (e.g., \fg{fg:rt_phia_x}).
Over a positive flux eddy, $\muC\sbi(s)>0$ 
represents local coherency of instantaneous flux from
right to left across $C$.
The sign and direction reverse over a negative flux eddy.
Therefore, the dynamic roots of $\muC\sbi(s)$ (\fg{fg:rt_phia_s}) can be
\fginsrt{fg:rt_phia_s}
traced back to dynamic eddies in $\chi\sbi(\vx)$  (\fg{fg:rt_psia})
through flux eddies in $\phi\sbi(\vx)$ (\fg{fg:rt_phia_x}).
The characteristic length-scale $\SCi$ of $\muC\sbi(s)$ is
defined by typical width of flux eddies 
in a unit of flight time by a reference trajectory along $C$.
Given $\phi\sbi(\vx)$, $\muC\sbi(s)$ depends critically on $C$
as shown for $\Ca$ and $\Cs$ of \model\ in \fg{fg:rt_phia_s} .
The two main controlling factors for $C$ 
are listed below.
$C$ may consist of more than one segment with different factors.

\begin{enumerate}
\item {\it Geographic location}.

The geographic location of $C$ with respect to flux eddies can lead to the
spatial phase conditions for $\muC\sbi(s)$ and $\muC\sbip(s)$:
counterparts of the temporal phase conditions for $f\sbi(t)$ and
$f\sbip(t)$  as in \eq{eq:ft} and \eq{eq:fiip}.

For a single mode $i$, if $C$ goes through the middle of flux eddies
in $\phi\sbi(\vx)$, then $\muC\sbi(s)$ may be regular and have
the spatial phase condition:
\bea
\muC\sbi(s)&\approx&-\muC\sbi(s+\SCi)~. \label{eq:muf_mu}
\eea
(left panels of \fg{fg:rt_phia_s} in comparison with  \fg{fg:rt_phia_x}).
In contrast, if $C$ runs near the edge of the flux eddies,
then $\muC\sbi(s)$ may be irregular
(right panels of \fg{fg:rt_phia_s}).

For a \pair\ $[i,i+1]$, if $C$ goes along (a part of) a flux alignment
curve, then $\muC\sbi(s)$ and $\muC\sbip(s)$ 
may be regular and have a similar structure over the segment.
However, there is a phase shift between them due to the staggered
centers of the flux eddies in $\phi\sbi(\vx)$ and $\phi\sbip(\vx)$.
Such a phase shift can be one of the following:
\bea\label{eq:ptype}
\muC\sbi(s)&\approx&\left\{
\begin{array}{ll}
\muC\sbip(s+\frac{\SCi}{2})~: \qquad & \mbox{phase type (i)}~, \\
\muC\sbip(s-\frac{\SCi}{2})~:        & \mbox{phase type (ii)}~.
\end{array} \right.
\eea
Phase type (i) corresponds to downstream propagation of the flux
eddies in $\phi\sbiip(\vx,t)$ (\sec{sec:flx2et_global}).
For \model, all \RTs\
belong to this  type along $\Ca$ (left panels of \fg{fg:rt_phia_s}). 
For phase type (ii), the flux eddies propagate upstream
against the reference flow,
as for the additional case discussed in Sections \ref{sec:model}
and \ref{sec:flx2et_global}.
In contrast, if $C$ has little association with a
flux alignment curve like $\Cs$ of \model,
then $\muC\sbi(s)$ and $\muC\sbip(s)$ are
unlikely to satisfy either spatial phase conditions
\eq{eq:ptype}.

\item {\it Kinematic type}.

If $C$ is either a semi- or bi-infinite boundary of transport associated
with an invariant manifold or separatrix like $\Cs$  (\sec{sec:et}), then
$\muC\sbi(s)$ must decay exponentially to zero as $s$ approaches
the corresponding DHT. 
Therefore, flux eddies must concentrate over some
bounded segments of $s$ away from the DHT.
In contrast, if $C$ has no DHT at the end points like $\Ca$,
then the flux eddies  may distribute along the entire segment of
$C$, 
depending on the geographic location of $C$ with
respect to flux eddies in $\phi\sbi(\vx)$.
If $C$ is a closed curve associated with a
periodic orbit of period $T\spC$, then $\muC\sbi(s)$ is 
also periodic.

\end{enumerate}

Having described $\muC\sbi(s)$ and  $\muC\sbip(s)$,
we consider $\muCa\sbi(s,t)$ and $\muCa\sbiip(s,t)$ 
as representative of evolving flux eddies extracted along $C$.
As a single mode $i$, individual flux eddies act like stationary
pistons that  synchronously push flux across $C$.
One push in each direction takes $\Tui$.
The non-dimensional characteristic scale ratio for the flux eddies is:
\bea\label{eq:gCi}
\GCi&\equiv&\frac{\SCi}{\Tui}~.
\eea 
As a \pair, $\muCa\sbiip(s,t)$ significantly depends on the
geographic location of $C$.
If $C$ goes along (a part of) a flux alignment curve of 
phase type (i), 
then the flux eddies act like sliding pistons along $C$.
Because they slide a distance $2\SCi$ in a unit of flight time
over $2\Tui$,
the non-dimensional phase speed is $\GCi$ in $\muCa\sbiip(s,t)$,
which relates to dimensional phase speed $b\spp\sbiip$ in
$\phi\sbiip(\vx,t)$. 
For phase type (ii), the non-dimensional phase speed is $-\GCi$.

\subsection{Spatio-temporal interaction} 
\label{sec:flx2et_et}

Having described $\muCa\sbi(s,t)$ and $\muCa\sbiip(s,t)$,
we identify the signature of dynamic variability in TIME
using the graphical approach.
Whenever possible, we use the subscript 
$\{\cdot\}\sbd$ to represent mode $i$, $i+1$  or pair
$[i,i+1]$ from here on.  
Using \eq{eq:m}, \eq{eq:da} and \eq{eq:muca}, 
the corresponding TIME  functions are: 
\bse\label{eq:etstm}
\bea
\mC\sbd(s,t;s_{a}:s_{b},t_{0}:t_{1})
&=&\int_{t_{0}}^{t_{1}} H(s-t+\tau;s_{a}:s_{b})
         \mua^{C,q}\sbd(s-t+\tau,\tau) d\tau~;
        \label{eq:etstm_m} \\
\aC\sbd(s,t;s_{a}:s_{b},t_{0}:t_{1})
&=&e(s:s_{a})\int_{t_{0}}^{t_{1}} H(s-t+\tau;s_{a}:s_{b})
        \mua^{C,e}\sbd(s-t+\tau,\tau)d\tau ~,
        \label{eq:etstm_a} 
\eea
and $\rC\sbd(\lC(s),t;s_{a}:s_{b},t_{0}:t_{1})=
        \aC\sbd(s,t;s_{a}:s_{b},t_{0}:t_{1})
        /||\vum(\vxCm(s))||$. 
\ese
For simplicity, we focus on 
$\aC\sbd(s,t;s_{a}:s_{b},t_{0}:t_{1})$
with $\mua^{C,e}\sbd(s,t)=\muCa\sbd(s,t)$,
which concerns
incompressible particle transport.
A full description of the TIME functions can be given
by replacing $\muCa\sbd(s,t)$ with $\mua^{C,q}\sbd(s,t)$ for
flow property $q$, or by explicitly including $e(s_{a}:s)$
in $\mua^{C,e}\sbd(s,t)$ for compressible particles.
It is worth adding that $\mua^{C,q}\sbd(s,t)$ and
$\mua^{C,e}\sbd(s,t)$ are roughly proportional to $\muCa\sbd(s,t)$ in
large-scale planetary flows.

The graphical approach examines the
integration of $\muCa\sbd(s-t+\tau,\tau)$ along a 
reference trajectory $(s-t+\tau,\tau)$ 
(Figures \ref{fg:rt_must_axis},
\fginsrt{fg:rt_must_axis}
\ref{fg:rt_must_cellu} and  
\fginsrt{fg:rt_must_cellu}).
In the $(s,t)$ space,  any reference trajectory is a diagonal line
with the unit phase speed $\ddrv{}{t}s=1$.
As a single mode, zero-$\muCa\sbd(s,t)$ contours divide the
$(s,t)$ space vertically for flux eddies with width $\SCi$, and
horizontally for phases of $f\sbi(t)$ with interval $\Tui$.
Individual ``boxes'' defined by these horizontal and vertical contours
have an aspect ratio $1/\GCi$ and represent spatial-temporal coherency
in the instantaneous flux along $C$.
At $\GCi=1$, a reference trajectory cuts across a box
diagonally: the flux mode is resonant with the reference flow over the 
eddies.
As a \pair\ for phase type (i),
the zero-contours with a positive slope $1/\GCi$ divide 
$\muCa\sbiip(s,t)$ into slices.
Each slice corresponds to a flux eddy
which propagates faster than a reference trajectory for $\GCi>1$, 
together at $\GCi=1$, and slower for $\GCi<1$.
For phase type (ii), the slope is negative $-1/\GCi$
and a reference trajectory always faces flux eddies head on.
If $C$ has little association with any flux alignment
curve, then the zero-contours may divide
$\muCa\sbiip(s,t)$ into a number of irregular regions
(right panel of \fg{fg:rt_must_cellu} for $\Cs$).

From a Lagrangian view point, integration \eq{eq:etstm} assumes
spatial-temporal interaction by moving with a reference 
trajectory.
The role of variability is determined by how the reference trajectory
encounters flux eddies along $C$.
From  the view of the mean-eddy interaction by the variability, it is 
determined by how boxes and slices distribute 
diagonally in the domain of integration $(s_{a}:s_{b},t_{0}:t_{1})$.
The rectangles surrounded by two vertical lines and two horizontal
lines in the right bottom panel of \fg{fg:rt_must_axis} 
(also left panel of \fg{fg:rt22rv_axis}) exemplify
a fixed domain of integration concerning definite TIME
(\sec{sec:et_mech}). 
If we wish to compute TIME associated with 
a specific event of flow dynamics, a more geometrically complex
domain can be judiciously chosen.
For finite-time TIME, the domain increases as
$t$ progresses.

To highlight the spatio-temporal interaction in \eq{eq:etstm}, 
we introduce spatial and temporal integrations of $\muCa\sbi(s,t)$
not associated with transport directly.
Spatial integration at time $t$ is 
$\int_{s_{a}}^{s_{b}}\muCa\sbi(s,t)ds=F\spC\sbi f\sbi(t)$,
where 
\bea\label{eq:Mm_F}
F\spC\sbi\equiv\sigma\sbi\int_{s_{a}}^{s_{b}} \muC\sbi(s) ds~:
\eea
is along a horizontal line and measures the spatial bias (\tbl{tbl:rt_Ca}).
If it is nonzero, then net instantaneous flux
oscillates with $f\sbi(t)$ and may lead to biased distribution of
signed pseudo-lobes.
\tblinsrt{tbl:rt_Ca}
In contrast, temporal integration is along a vertical line.
Over the entire time interval  $t\in[t_{-},t_{+}]$,
it must be zero at any $s$, i.e., 
$\int_{t_{-}}^{t_{+}} \muCa\sbi(s,t) dt=0$.

\section{Graphical approach}
\label{sec:graphic}

Using \model, we demonstrate the graphical approach for individual cases.
Given $\aC\sbd(s,t;s_{a}:s_{b},t_{0}:t_{1})$,
spatial coherency and temporal evolution of transport
is naturally described in terms of pseudo-PIPs and pseudo-lobes (\sec{sec:et}).
Some phenomena described below will be revisited in the next section
using the analytic approach.

\subsection{Finite-time TIME across the jet axis $\Ca$}
\label{sec:graphic_Ca}
We consider the finite-time TIME
$\aa\sbd(s,t)=\aa\sbd(s,t;s\spdgr:s,t\spdgr:t)$ for $\Ca$,
from time $t\spdgr$ up to the present $t$.
We use the spatial periodic condition
$\muaa\sbd(s,t)=\muaa\sbi(s+T\spa,t)$,
where $T\spa=1.260549$ is the period of $\Ca$.
As $t$ progresses, $\aa\sbd(s,t)$  changes because
$\muaa\sbd(s,t)$ keeps adding to the integration.
\fg{fg:rt_ms_axis_range} shows the evolution of pseudo-lobes as 20
envelopes of $\aa\sbd(s,t)$ taken at every $T\spa/2$.
\fginsrt{fg:rt_ms_axis_range}

We start with the third pair $[5,6]$
whose pseudo-lobes have larger width and amplitude than any other
pairs.
Neither mode 5 nor 6 has spatial bias due to 
the symmetry (\fg{fg:rt_phia_x}).
Mode 5 has a total of six flux eddies along $\Ca$.
The two opposite-signed ones at $s=0.3157$ and $0.9454$ are much
stronger than the others (\fg{fg:rt_phia_s}).
As they pulsate,
a diagonal line  can go through the boxes
associated with these two flux eddies when they have the same sign.
By considering only them,
the two opposite-signed pseudo-lobes emanate along $\Ca$ 
in $\aa\sbmd{5}(s,t)$.
It also leads to $\Gamma\spa\sbmd{5}\sim 1$:
the most efficient spatio-temporal interaction 
as we shall see in the next section.
In mode 6, the two pairs replace those two flux eddies in mode 5
(\fg{fg:rt_must_axis}).
This process can be also thought as $\Gamma\spa\sbmd{6}\sim 1$;
however,  it is not as efficient as that for mode 5.
Therefore, $\aa\sbmd{6}(s,t)$ has two opposite-signed pseudo-lobes with
smaller amplitude.
Given the temporal phase condition \eq{eq:fiip} and the 
spatial phase type (i) of \eq{eq:ptype}, modes 5 and 6 
cooperate to form two large pseudo-lobes in $\aa\sbmd{[5,6]}(s,t)$.

For any other pairs, one of the two modes has no spatial bias
and almost satisfies the spatial regularity,
i.e., 
$|F\sbi\spa|=0$  and  $S\spa\sbi=T\spa/2N\spp\sbi$ for $i=1,4,8$ and $9$
(\fg{fg:rt_must_axis} for mode 1).
The pseudo-lobes of these modes have a wave number 
$N\spp\sbi$ pattern.
Other modes have non-zero $|F\sbi\spa|$,
mainly due to the  same-signed flux eddies centered 
at $s=0.3157$ and $0.9454$.
This $|F\spa\sbi|$ effect leads to a
biased distribution of the signed pseudo-lobes.
It is more substantial for even modes, embraced by the 
phase of $f\sbi(t)$ starting from $t\spdgr(=0)$
until $\Tui$ when the phase reverses.
Accordingly, $\aa\sbmd{2}(s,t)$ is heavily dominated by negative
pseudo-lobes due to the spatial,
temporal, and spatio-temporal reinforcers as follows:
1) $|F\spa\sbmd{2}|$ is large because the two negative flux eddies
are much wider and stronger than the two positive ones;
2) the phases of $f\sbmd{2}(t)$ heighten the $|F\spa\sbmd{2}|$ effect;
3) the small characteristic ratio $\Gamma\spa\sbmd{2}\ll 1$ 
allows the $|F\spa\sbmd{2}|$ effect to spread over $\Ca$ along 
all diagonal lines 
(\fg{fg:rt_must_axis} for mode 2).
Biased pseudo-lobes are also observed in $\aa\sbmd{10}(s,t)$,
but not as significant as in $\aa\sbmd{2}(s,t)$,
because the spatial and spatio-temporal reinforcers are weaker
for mode 10.
Odd modes show little  bias and have 
$N\spp\sbi$ pairs of signed pseudo-lobes in $\aa\sbmd{i}(s,t)$.
As a \pair,
negative pseudo-lobes almost always dominate $\aa\sbmd{[1,2]}(s,t)$.
The pseudo-lobes in
$\aa\sbmd{[3,4]}(s,t)$, $\aa\sbmd{[7,8]}(s,t)$ and 
$\aa\sbmd{[9,10]}(s,t)$ have a slight bias in their
wave number $N\spp\sbtnn$ pattern.

How can these results be interpreted as atmospheric transport
phenomena?
We recall that the $n$-th \RT\ $(\knb,\lnb)$
generates flux eddies along the  
reference jet as it propagates (straight) eastward with phase speed 
$b\spu\sbmd{[2n-1,2n]}$ along $\lnb$ lines.
The flux eddies so generated 
intensify when passing the trough and ridge where the jet flows
faster.
The phase speed $b\spp\sbmd{[2n-1,2n]}$ is slower than 
$b\spu\sbmd{[2n-1,2n]}$.
All waves have positive characteristic ratios $\Gamma\spa\sbtnn<1$.
Therefore, a reference trajectory passes flux eddies from behind.
  
For the third Rossby traveling wave $(k\sbwv{3},l\sbwv{3})=(2,2)$,
this process leads to synchronized bursts of opposite-signed
flux eddies over the trough and ridge that change sign 
every $T\spu\sbmd{5}$.
The time interval between consecutive
bursts approximately equals the flight-time of a reference particle from
the trough to ridge, i.e., 
$T\spu\sbmd{5}\sim T\spa/2$.
Therefore, a particle can be almost always pushed in the same
direction by every burst. 
Thus, the envelope of transport has large amplitude.
The wave number 1 pattern mainly emanates opposite-signed
bursts at the ridge and trough.
This process can be thought of as resonance between the reference
jet and the wave.

For the first \RT\ $(k\sbwv{1},l\sbwv{1})=(1,2)$,
the bursts have the same sign in mode 2, leading to a strong
southward spatial bias $F\spa\sbmd{2}$.
The positive phase of $f\sbmd{2}(t)$ supports this effect
starting from $t\spdgr=0$ up to $\Tui$ until the phase reverses.
The temporal variability of this wave is slow enough 
that kinematic advection by the underlying reference flow can
spread the spatial bias effect all along $\Ca$.
Thus, this wave strongly favors southward finite-time transport
starting at $t\spdgr=0$.
If starting at $t\spdgr=T\spu\sbmd{1}$, then it favors northward.
The envelopes of transport show a shadow of
a wave number $N\spp\sbmd{1}=2$ pattern.

The fifth \RT\ $(k\sbwv{5},l\sbwv{5})=(3,2)$ has a
northward spatial bias $F\spa\sbmd{10}$.
Its effect is not as distinguished as $F\spa\sbmd{2}$
because the bursts are weaker and the temporal variability is too
fast to spread the spatial bias.
Thus, the envelopes of transport almost always
sustain a wave number $N\spp\sbmd{10}=3$ pattern with slight
northward bias.

The second and fourth \RTs\
$(k\sbwv{2},l\sbwv{2})=(2,1)$ and $(k\sbwv{4},l\sbwv{4})=(2,3)$  
have little bias in the distribution of signed pseudo-lobes,
because $f\sbi(t)$ dissipates the spatial bias.
The envelopes of transport have the wave number 
$N\spp\sbmd{3}=2$ and $N\spp\sbmd{7}=3$ patterns, respectively.

For all waves, the area of pseudo-lobes fluctuates
as they propagate along $\Ca$.
Surprisingly, the fluctuation frequency and propagation phase speed 
have no direct tie to the corresponding flux eddies
or reference flow.
Using an analytic approach, we explain these phenomena
in \sec{sec:analytic_finite}.

\subsection{Definite TIME across the separatrix
$\Cs$}
\label{sec:graphic_Cs}
We consider definite TIME
$\as\sbd(s,t)=\as\sbd(s,t;-\infty:\infty,-\infty:\infty)$
across $\Cs$.
We refer to  the two segments of $\Cs$ as the
``western'' ($s<0$) and ``eastern'' ($s>0$) regions
with respect to the trough ($s=0$).
Due to the particular choice for the domain of integration 
along the separatrix,
pseudo-lobes of $\as\sbd(s,t)$ relate to lobes of
Lagrangian transport theory up to the leading order (this is explained in \cite{IW_TIMEI}).
Because $\Cs$ is away from the flux alignment curve,
any mode has an irregular distribution of flux eddies
(\fg{fg:rt_phia_s}).
Nonetheless,
$\as\sbd(s,t)$ shows some remarkable similarities. 
The pseudo-PIP sequence has a constant separation
distance $\Tui$ in a unit of flight time.
The pseudo-lobe sequence spreads homogeneously along $\Cs$.
Because the sequence propagates with the reference flow without
deformation in the $(s,a)$ space,
it suffices to consider one snap shot of  $\as\sbd(s,t)$ 
(\fg{fg:rt_ms_cellu_t5}).
\fginsrt{fg:rt_ms_cellu_t5}
These are general properties of definite TIME
and will be explained in \sec{sec:analytic_definite}.
Due to the (anti-)symmetry in $\mu\sps\sbi(s)$ and $f\sbi(t)$,
any $\as\sbd(s,t)$ share the same pseudo-PIP sequence.

For example, mode 5 has two flux eddies 
in $\mu\sps\sbmd{5}(s)$ over both the western and eastern regions 
(\fg{fg:rt_phia_s}).
The second and third flux eddies around the trough are positive.
They are stronger than the first and fourth ones that are negative.
The characteristic ratio over the positive ones are slightly
larger than 0.5 in $\musa\sbmd{5}(s,t)$ (\fg{fg:rt_must_cellu}).
Hence, a diagonal line can go through just about all boxes of 
spatio-temporal coherency  while they have the same sign. 
For a given $\mu\sps\sbmd{5}(s)$, $f\sbmd{5}(t)$ has an effective
$T\spu\sbmd{5}$ that enhances spatio-temporal interaction in
$\musa\sbmd{5}(s,t)$: 
all flux eddies may favor inflating the pseudo-lobes of
$\as\sbmd{5}(s,t)$. 
In contrast,  four flux eddies of the complementary mode 6
alternate the signs in $\mu\sps\sbmd{6}(s)$ and 
have relatively small characteristic ratios in $\musa\sbmd{6}(s,t)$.
Contributions from these pulsating flux eddies 
tend to hinder each other
along the diagonal line, resulting in the flat pseudo-lobes
of  $\as\sbmd{6}(s,t)$.
Mode 5 rules $\as\sbmd{[5,6]}(s,t)$ when the two modes are combined
(\fg{fg:rt_ms_cellu_t5}).

For the $n$-th \RT\ as a pair $[2n-1,2n]$,
downstream propagation of flux eddies  in
$\musa\sbmd{[2n-1,2n]}(s,t)$ 
can be irregular and discontinuous (\fg{fg:tr_must_cellu}).
\fginsrt{fg:tr_must_cellu}
For the first, third, and fifth \RTs,
propagation splits into the two segments  
because the flux eddy along the alignment curve does not reach $\Cs$
at the trough ($x=1/2$) 
in $\phi\sbmd{2}(\vx)$, $\phi\sbmd{7}(\vx)$ and $\phi\sbmd{10}(\vx)$,
respectively. 
This leads to the propagation that looks as if the flux eddies 
disappear at the trough ($s=0$) and reappear in the downstream
with a reversed sign. 
For the second \RT, propagation in
$\musa\sbmd{[3,4]}(s,t)$ is irregular but continuous 
because all flux eddies in $\phi\sbmd{3(\vx)}$ and $\phi\sbmd{4}(\vx)$
reach $\Cs$ for $0<x<1$.
The fourth \RT\ also has a continuous propagation in 
$\musa\sbmd{[7,8]}(s,t)$,
although $\Cs$ skips a couple of flux eddies in $\phi\sbmd{7}(\vx)$ 
and $\phi\sbmd{8}(\vx)$ along the flux alignment curve.
For these second and fourth \RTs, propagating flux eddies in
$\musa\sbmd{[2n-1,2n]}(s,t)$ intensify when passing the trough.

Using the graphical approach, we examine the role of \RTs\ in transport
with a focus on a pseudo-lobe.
We show what the positive pseudo-lobe 
${\cal L}\spstr={\cal L}\spu\sbmd{[2n-1,2n],0:1}(t\spstr)$
closest to the trough along $\Cs$ at time $t\spstr=5$ is made of.
The two pseudo-PIPs that define ${\cal L}\spstr$ are shown in
\fg{fg:rt_ms_cellu_t5} by the circles.
We denote the upstream  pseudo-PIP by
$s\spu\sbmd{[2n-1,2n],0}(t\spstr)=s\spstr$.
Then the downstream one is
$s\spu\sbmd{[2n-1,2n],1}(t\spstr)=s\spstr+T\spu\sbmd{2n-1}$.

The two diagonal lines  in \fg{fg:tr_must_cellu} go through the 
two pseudo-PIPs.
Because the definite TIME $a\spu\sbmd{[2n-1,2n]}(s,t)$
propagates with the reference flow, they also coincide with the
trajectories of the two pseudo-PIPs.
The region  bounded by these diagonal lines 
shows the makeup of ${\cal L}\spstr$.
The propagation phase speed of flux eddies is slower than
a reference trajectory for any wave,  i.e., $\Gamma\sps\sbmd{2n-1}<1$.

For the first, third, and fifth \RTs, the region contains
four propagating flux eddies.
Over the western region, 
a positive follows a negative.
The order reverses over the eastern region.
The maximum amplitude of ${\cal L}\spstr$ occurs at 
$s\spstr+T\spu\sbmd{2n-1}/2$.
The corresponding reference trajectory 
moves through the positive flux eddies over the
western region, reaches the trough at the same time as
the head of the same flux eddy, and moves into another positive flux
eddy over the eastern region.
Therefore, $a\sps(s\spstr+T\spu\sbmd{2n-1}/2,t\spstr)$ has
equal and positive contributions from  both regions.
Such spatio-temporal interaction can be more efficient for smaller
$\Gamma\sps\sbmd{2n-1}$ if $\Gamma\sps\sbmd{2n-1}<1$.
Therefore, the first \RT\ is most efficient among the three due to the
large $\Tu\sbmd{1}$.
At $\Gamma\sps\sbmd{2n-1}\approx 1$,
contribution from either region becomes zero.
(compare $\mu\sps\sbmd{[1,2]}(s,t)$ and $\mu\sps\sbmd{[9.10]}(s,t)$
in \fg {fg:tr_must_cellu}).

In the upstream portion of ${\cal L}\spstr$,
a negative contribution develops from the trough due to the 
second propagating flux eddy over the eastern region
and it propagates upstream.
At the upstream pseudo-PIP $s\spstr$, this negative contribution
balances the positive contribution from the western region.
Conversely, in the downstream portion of ${\cal L}\spstr$,
negative contribution develop from the trough over the western
region and propagate downstream.
At the downstream pseudo-PIP $s\spstr+T\spu\sbmd{2n-1}$,
it balances the positive contribution from the eastern region.

By tracing its dynamic roots in $\psia\sbmd{[2n-1,2n]}(\vx,t)$,
we find that the reference trajectory
$(s\spstr+T\spu\sbmd{2n-1}/2-t\spstr+\tau,\tau)$
for the maximum pseudo-lobe amplitude 
goes through the trough at the same time as the center of a
positive dynamic eddy  in the south reaches the trough 
over the recirculating cell.
In addition, 
$(s\spstr+T\spu\sbmd{2n-1}-t\spstr+\tau,\tau)$ for the
downstream pseudo-PIP and 
the head of that dynamic eddy arrive at the
trough  simultaneously;
$(s\spstr-t\spstr+\tau,\tau)$
for the upstream pseudo-PIP and the tail of the dynamic eddy 
go through the trough  simultaneously.

Accordingly, the pseudo-lobe sequence of these \RTs\ appears to 
evolve as if they move with the dynamic eddies in $\vx$.
Note that the pseudo-lobe sequence itself propagates with the
reference flow along $\Cs$.

The role of variability completely differs for the second  and fourth \RTs.
The two diagonal lines associated with ${\cal L}\spstr$
mainly contain  a  positive propagating flux eddy. 
Due to $\Gamma\sps\sbmd{2n-1}<1$, the western region includes
a negative propagating flux eddy that succeeds the positive.
Over the eastern region, a negative one precedes.

The center of the main positive flux eddy and 
the reference trajectory
$(s\spstr+T\spu\sbmd{2n-1}/2-t\spstr+\tau,\tau)$ going through the
center of ${\cal L}\spstr$ reaches the trough simultaneously.
Therefore the trajectory receives the maximum contribution
from it.

Clearly,  the maximum  amplitude of ${\cal L}\spstr$ occurs along the
reference trajectory that goes though the trough at the same time
as the center of the positive flux eddy does.
In the upstream portion of ${\cal L}\spstr$,
a negative contribution develops from the upstream of $\Cs$
due to the succeeding negative flux eddy over the western region.
In contrast, the downstream portion of ${\cal L}\spstr$ develops
a negative contribution from the downstream of $\Cs$ 
due to the preceding negative flux eddy over the eastern regions.

Such spatio-temporal interaction along the diagonal line is most
efficient at $\Gamma\spu\sbmd{2n-1}\approx 1$, unlike the three
\RTs\ mentioned above.
Moreover, we find that the spatio-temporal interaction
for the fourth \RT\ is inefficient, because intensifications of
the positive flux eddy off the trough (around $|s|\approx 0.3$)
are wasted along the two diagonal lines for the pseudo-PIPs.
If $T\spu\sbmd{7}$ decreases to increase $\Gamma\sps\sbmd{7}$ to near 1,
then these intensifications can significantly help enhance $a\sps(s,t\spstr)$ 
around $s\sim s\spstr+T\spu\sbmd{2n-1}/2$.

By tracing the dynamic root of the spatio-temporal interaction
for the second \RT, we find that the center of a positive dynamic eddy
in $\psia\sbmd{[3,4]}(\vx,t)$ and the reference trajectory associated
downstream pseudo-PIP go though the trough simultaneously.
After $T\spu\sbmd{3}$, the center of the succeeding positive dynamic
eddy and the reference trajectory of the upstream pseudo-PIP
go though the trough together.
Accordingly, the evolution of the pseudo-lobe sequence appears to succeed
the dynamic eddies with a $T\spu\sbmd{3}/2$ time-lag.

For the fourth \RT, two southern dynamic eddies in each
column act together.
The evolution of the pseudo-lobe sequence succeeds the dynamic eddies
in the middle row with a $T\spu\sbmd{7}/2$ time-lag.

Definite TIME due to all five \RTs\ is shown in 
\fg{fg:tr_must_cellu_t5}.
The graphical approach can be used in two ways.
Given $a\spu(s,t)$, 
comparing $\mua\sbmd{[2n-1,n]}\spu(s,t)$ to $\mua\spu(s,t)$
identify the variability which is responsible for 
individual pseudo-lobes.
Inversely, given individual $\mua\sbmd{[2n-1,n]}\spu(s,t)$,
we obtain how they collaborate to produce net $a\spu(s,t)$.
Finally, we note that the graphical approach can be used 
to identify role of variability when $f\sbtnn(t)$ is
irregular, i.e., $T\spu\sbtnn$ varies in time.

\fginsrt{fg:tr_must_cellu_t5}
%

\section{Analytical approach} 
\label{sec:analytic}

Motivated by \sec{sec:graphic},
we now explore the underlying properties of finite-time and definite
TIME that are  relevant to large-scale geophysical flows.
We highlight the results of the analytical approach in this section
while details are provided in \appndx{appndx:analytic}.

\subsection{Finite-time TIME}
\label{sec:analytic_finite}

To identify  the basic properties of finite-time transport
$\aC\sbd(s,t)=\aC\sbd(s,t:s-t+t\spdgr:s,t-t\spdgr:t)$,
we consider an idealized case:
$\muCa\sbd(s,t)$ arising from regularly
distributed  flux eddies 
(see upper panels of \fg{fg:rt_must_axis}). 
If $C$ is periodic with period $T\spC$ and $\muC\sbi(s)$ has
$2N\spp\sbi$  flux eddies like $\Ca$, then $\SCi=T\spC/2N\spp\sbi$
gives $\GCi=T\spC/(2N\spp\sbi\Tui)$. 
First, we revisit the graphical approach
$\aC\sbd(s,t)$.
This time, we focus on a single diagonal line  of integration where the
evaluation point $s=s\spdgr-t\spdgr+t$ moves downstream as $t$ increases.

\renewcommand{\theenumi}{\alph{enumi}}
\begin{itemize}
\item Mode $i$:
The corresponding $\muCa\sbi(s,t)$ has uniform boxes of a size
$\SCi\times\Tui$ (\fg{fg:rt_must_axis}).
Integration \eq{eq:etstm_m} may be most efficient for $\GCi\approx 1$,
because $\aC\sbi(s,t)$ can increase monotonically if the diagonal line
of integration goes through one box to another without ever changing the sign of
$\muCa\sbi(s\spdgr-t\spdgr+t,t)$ as $t$ increases.
In contrast, for $\GCi\not\approx 1$, $\aC\sbi(s,t)$ fluctuates because
$\muCa\sbi(s\spdgr-t\spdgr+t,t)$ changes  sign regularly
by running through one box to another.
For $\GCi\gg 1$, 
$\aC\sbi(s,t)$ oscillates with a period roughly $2\Tui$
because the diagonal line goes through horizontally-thin boxes  in
the $(s,t)$ space.
In contrast, for $\GCi\ll 1$, $\aC\sbi(s,t)$ oscillates with a period
roughly $2\SCi$. 
\item Pair $[i,i+1]$:
The corresponding $\muCa\sbiip(s,t)$ has slanting contours in the $(s,t)$
as shown in right bottom panel of \fg{fg:rt_must_axis} for phase type
(i) and left panel of \fg{fg:rt22rv_axis} for (ii).
It is  clear  that $\aC\sbiip(s,t)$ for phase type (ii)
oscillates faster
because $\muCa\sbiip(s,t)$ changes  sign 
more frequently along the diagonal line.
\end{itemize}
\fginsrt{fg:rt22rv_axis}

Next, we use 
the analytical approach on the idealized models
to quantify the spatio-temporal interaction as a function of $\GCi$
(\appndx{appndx:analytic_1}, Model case 1.1).
Highlights of the phenomena follow.

\begin{itemize}
\item Mode $i$: Finite-time TIME can be represented as a sum
of low- and high-frequency responses to pulsating flux
eddies whose amplitude and period depend on $\SCi/(1-\GCi)$ and
$\SCi/(1+\GCi)$, respectively \eq{eq:muCa}.
Based on amplitude, the low-frequency tends to overcome the
high-frequency response; especially for $\GCi\approx 1$.
At $\GCi=1$,
the low-frequency response becomes resonant and exhibits linear growth as in
\eq{eq:muCa_b}.
\tbl{tbl:rt_Ca} summarizes
the predicted values of the amplitude and period along $\Ca$.
\item Pair $[i,i+1]$: The response is restricted to
low-frequency for phase type (i) and high-frequency for phase
type (ii), confirming the results using the graphical approach.
\end{itemize}

We see from \tbl{tbl:rt} that 
$\GCi$ is fairly small with respect to 1
for any \RT.
This means that, if the wave travels faster,
then more TIME can be expected.
If we use $N\spp\sbmd{3}=1$ (\sec{sec:graphic}) for pair $[3,4]$, then
$\Gamma\spC\sbmd{3}\approx 1$ (\tbl{tbl:rt_Ca}) leads to a large
values of $\SCi/(1-\GCi)$.
This would explain the extremely large pseudo-lobes
when compared to other pairs.

Using the spatial phase conditions \eq{eq:muf_mu} and \eq{eq:ptype}, 
\bea\label{eq:a_cat1}
\aC\sbd(s,t)\approx-\aC\sbd(s-\SCi,t)~
\eea
holds independent  of  the  temporal components $f\sbi(t)$ and
$f\sbip(t)$. 
This confirms the observations made in \sec{sec:graphic}
that $\aC\sbd(s,t)$ along $\Ca$ may have the same wave
number as the flux eddies for $|F\spC\sbi|\neq 0$ is nonzero.
Therefore, the separation distance of the  pseudo-PIP sequence
is a constant 
$\triangle s\spC\sbmd{\bcdt,j:j+1}=\SCi$.
At time $t$, the corresponding
pseudo-lobe sequence has a homogeneous configuration with alternating
sign along $C$.

The  analytical approach in \appndx{appndx:analytic_1} (Model case 1.2)
relates the geometry  and evolution of pseudo-lobes to 
the characteristic ratio $\GCi$. 
For simplicity, we summarize the two cases
of response for $\aC\sbiip(s,t)$.
The case  for $\aC\sbi(s,t)$ is given as a linear sum of the two responses.

\begin{itemize}
\item A low-frequency response for phase type (i) \eq{eq:pyi}:
A pseudo-lobe sequence
propagates downstream at a constant
phase speed $(1+\GCi)/2$ \eq{eq:pyi_pip}, which averages the 
phase speed of a reference trajectory and flux eddy propagation.
At $\GCi=1$ when the flux eddies propagate with the reference flow,
so too do all pseudo-lobes whose  signed area
$A\spC\sbmd{[i,i+1],j:j+1}(t)$ grows linearly in $t$ \eq{eq:pyi_A}.
For $\GCi\neq 1$, maximum amplitude of
$A\spC\sbmd{[i,i+1],j:j+1}(t)$ oscillates with the period
$4\SCi/(1-\GCi)$.
As $\GCi$ varies from 1, $A\spC\sbmd{[i,i+1],j:j+1}(t)$ decays like
$1/\GCi$.

\item A high-frequency response for phase type (ii) \eq{eq:pyii}:
The same formulae obtained for the low-frequency response hold by 
replacing $\GCi$ with $-\GCi$.
The main difference appears in propagation where all pseudo-lobes
travel downstream for $\GCi<1$ and upstream for $\GCi>1$.   
At $\GCi=1$, $\aC\sbmd{[i,i+1]}(s,t)$  behaves like standing waves 
with oscillation period $2\Tui$.
\end{itemize}

\tbl{tbl:rt_Ca} shows the predicted phase speed of
the pseudo-lobes for \example.

\subsection{Definite TIME}
\label{sec:analytic_definite}
Here we explore general properties associated with definite TIME
$\aC\sbd(s,t)=\aC\sbd(s,t;s_{a}:s_{b},t_{0}:t_{1})$ for
a fixed  $(s_{a}:s_{b},t_{0}:t_{1})$,
where $C$ is not necessarily a separatrix.
Substituting temporal phase conditions \eq{eq:ft} and \eq{eq:fiip}
into \eq{eq:etstm_a} gives two useful relations:
\bse\label{eq:aC_c2}
\bea
\aC\sbd(s,t)&\approx&-\aC\sbd(s-\Tui,t)~, 
\label{eq:aC_c2_1}\\
\aC\sbd(s,t)&\approx&
   \aC\sbd(s+\triangle t,t+\triangle t)~.\label{eq:aC_c2_2}
\eea
\ese
The first relation means that the separation distance
$\triangle s\spC\sbmd{\bcdt,j:j+1}$ of the pseudo-PIP sequence
is a constant  $\Tui$ and 
the pseudo-lobe sequence has a homogeneous configuration 
with alternating signs in the $(s,a)$ space.
This result can be confirmed graphically along the two diagonal lines
of integration separated by $\Tui$ in $t$ (\fg{fg:rt_must_cellu}).
The second relation means that
the pseudo-PIP and pseudo-lobe sequences propagate
with the reference flow  without deformation  in the $(s,a)$ space.
This invariance can be confirmed graphically along any diagonal line.

Motivated by 
\sec{sec:graphic_Cs} using the graphical approach,
we consider the following two cases.
The first case focuses on  contributions from individual flux eddies
in $\muC\sbi(s)$.
This is useful to understand not only the role of individual flux
eddies but also how they are combined in the total transport.
The second case considers the spectral response of the spatial component
$\muC\sbi(s)$ to the temporal PC $f\sbi(t)$.
This helps us understand the effect of having more than one spectral component in $f\sbi(t)$
or sensitivity with respect to temporal frequency.
Details of both analyses are given in \appndx{appndx:analytic_2}
for both cases.

We first consider the contribution of $\aC\sbmd{\bcdt.p}(s,t)$
from the $p$-th flux eddy,
where the subscript $\{\cdot\}\sbmd{\bcdt.p}$ corresponds to
the $p$-th flux eddy of variability $\bcdt$.
The relations in \eq{eq:aC_c2} hold for $\aC\sbmd{\bcdt.p}(s,t)$
by adjusting $(s_{a}:s_{b},t_{0}:t_{1})$ to a corresponding 
vertical column in $\muCa\sbd(s,t)$ 
\eq{eq:sasb_p}.
The analytical approach based on a highly idealized model in
\appndx{appndx:analytic_2} (Model case 2.1) illuminates the 
dependence on $\GCidp$, as briefly summarized below.

\begin{itemize}
\item Mode $i$:
Definite TIME $\aC\sbidp(s,t)$ splits into two
elements: amplitude and normalized configuration
\eq{eq:mCi_st}.
The normalized configuration is based on $f\sbi(s-t)$,
confirming the constant separation distance 
$\Tui$ of the pseudo-PIP sequence.
The amplitude $M\spC\sbidp$ has a maximum at  $\GCidp=1$ and decays
like $\SCi/\{1-(\GCidp)^{2}\}$ with fluctuations \eq{eq:mCi_M}.
Since the fluctuation is due to cancelation along the diagonal line by
frequent sign changes of $f\sbi(t)$,
for an even integer $\GCi$,  the spatio-temporal interaction is subharmonic
and cancels the net transport completely.

\item Pair $[i,i+1]$: For the complementary mode of the pair,
$\aC\sbipdp(s,t)$ also splits into the two elements.
The  normalized
configuration is based on $f\sbi(s-t)$, but there
is a phase shift $\pm 2\theta\spC\sbiipdp$ from $\aC\sbidp(s,t)$
where $+$ and $-$ for phase type
(i) and (ii) of \eq{eq:ptype}, respectively.
As a pair, amplitude $M\spC\sbmd{[i,i+1].p}$ 
 has a fluctuating multiplication factor 
to $M\spC\sbidp$,
 while the normalized configuration has 
a phase shift $\pm\theta\spC\sbiipdp$ from $\aC\sbiipdp(s,t)$
At $\GCidp=1$, phase type (i) has the maximum enhancement 
$\aC\sbmd{[i,i+1].p}(s,t)=2\aC\sbidp(s,t)$ 
while phase type (ii) results in a complete suppression
$\aC\sbmd{[i,i+1].p}(s,t)=0$.
Therefore, $\GCidp$ governs how the two modes may incorporate
 each other in forming $\aC\sbmd{[i,i+1].p}(s,t)$.
\end{itemize}

Given contributions from individual flux eddies, 
the total $\aC\sbd(s,t)$ can be given as a linear sum of
individual contributions
(\appndx{appndx:analytic_2}, Model case 2.2a).

The second is more general.
It considers the total $\aC\sbd(s,t)$ as a spectral response of the
flux spatial component $\muC\sbd(s)$ 
to the temporal variability $f(t)=\cos\pi t/T$ where $T$ is the
characteristic time scale
(\appndx{appndx:analytic_2}, Model case 2.2b).
It is easy to show that $\aC\sbd(s,t)$ consists of an amplitude
$M\spC\sbi$ and normalized configuration $f(s-t+\theta)$ where
$\theta-t$ is the phase lag between the temporal forcing and
geometry at a fixed $t$.
Since $\muC\sbi(s)$ is given without specifying $\SCi$,
the efficiency of the spatio-temporal  interaction is given as a function
of $1/T$, rather than $\GCi=\SCi/\Tui$  \eq{eq:msf_cs}.
By definition, the limit of the amplitude corresponds to the spatial
bias: $\lim_{1/T\to 0} M\spC\sbi=F\spC\sbi$ \eq{eq:Mm_F}.

\fg{fg:rt_Mcs}
\fginsrt{fg:rt_Mcs}
shows $M\spC\sbi$ as a function of $1/T$ for \model\ along $\Cs$.
It also gives the sensitivity of definite transport
with respect to the temporal component, given the 
spatial component of dynamic variability.
We confirm our observation in \sec{sec:graphic_Cs}
that the spatio-temporal interaction of the third \RT\ is efficient
and that it is not for the forth and fifth \RTs.
We also confirms that the forth \RT\ can be enhanced significantly by
decreasing $T$.
The first and second \RTs\ are also found to be efficient from the figure.

\section{Concluding remarks}
\label{sec:cncl}

By combining a spatio-temporal analysis for variability and the
geometrical method for transport, we have studied how variability 
affects transport of  fluid particles and flow properties 
across a stationary boundary curve defined kinematically by a streamline of
the mean flow. 
The Transport Induced by Mean-Eddy Interaction (TIME) theory
is  natural for this because it presents
transport as the integration of the spatio-temporal 
interaction between the steady reference flow
and the unsteady anomaly.
Through the spatially nonlinear interaction with the reference flow,
the dynamic mode $i$ in the velocity anomaly leads to the flux
mode $i$.
Therefore, global evolution of dynamic eddies and flux eddies in the
flow field can be spatially different.
Along the boundary curve $C$ across which we evaluate transport,
the characteristic length-scale $\SCi$  is determined by
the flux eddies  in a unit of flight-time, while the characteristic
time-scale $\Tui$  is given by the corresponding dynamic mode.
The non-dimensional characteristic ratio $\GCi=\SCi/\Tui$ rates 
as the most important parameter by providing
the spatio-temporal coherency of instantaneous flux.
At $\GCi=1$, the flux variability and  the reference flow dynamics are 
resonant.
When modes $i$ and $i+1$ are a dynamically coherent pair,
$\GCi$ may correspond to the propagation speed of flux eddies along $C$.

We have presented a graphical approach for individual cases and an
analytical approach for general properties of transport.
We focused on the two types of transport that may be most 
relevant to large-scale geophysical flows.
One concerns finite-time transport  starting from an initial time 
up to the present time.
For each mode, the transport consists of low- and high-frequency
responses to the pulsating flux eddies.
The low-frequency response relates to 
downstream propagation of flux eddies and high-frequency response
to upstream.
The pseudo-lobes of both responses have
the same width $\SCi$ as the flux eddies.
The propagation phase speed averages the
reference flow and flux eddies, i.e., envelopes of transport is 
associated neither with particle motion nor dynamics (or even flux) waves.
The amplitude and oscillation period of pseudo-lobes depend on
$\SCi/(1-\GCi)$ for low-frequency response and
$\SCi/(1+\GCi)$ for high-frequency response, respectively.
Hence, the low-frequency response tends to dominate
the high. 

The other type concerns definite transport.
The corresponding pseudo-lobes propagate with the reference 
flow without changing  area.
The width of the pseudo-lobes is determined by the characteristic
time-scale $\Tui$ as flight-time along $C$ and has nothing to do with
flux eddy structure.
The amplitude decays like
$\SCi/\{1-(\GCi)^2\}$ with fluctuation as $\GCi$ increases. 
Moreover, the pseudo-lobes appear to move with the
dynamic eddies with a possible phase-lag, although it propagates
downstream with the reference flow.

In the case  where $C$ is a separatrix, the spatial segment and temporal
interval 
can be chosen to be bi-infinite.
Such a definite transport is equivalent to the Melnikov function for
Lagrangian lobe dynamics (as shown in \cite{IW_TIMEI}).
Therefore, it can shed some light on the role of
variability for  Lagrangian particle transport as well.

By applying the graphical and analytical approaches to the kinematic model
of the large-scale atmospheric flow for a blocked flow,
we have examined the role of  \RTs\ in transport.
We have identified similarities and differences in their spatio-temporal
interaction.

Because of its flexibility, the framework presented here can be
modified and extended for  various types of transport studies.

\section*{Acknowledgements}

This research is supported by ONR Grant No.~ N00014-09-1-0418 and N00014-10-1-055 (KI),  ONR Grant No.~N00014-01-1-0769 (SW) and by
MINECO under the ICMAT Severo Ochoa project SEV-2011-0087. 

\appendix
\renewcommand{\theequation}{\Alph{section}.\arabic{equation}}
\setcounter{equation}{0}
\section{idealized model for TIME}
\label{appndx:analytic}
\subsection{Category 1: Finite-time TIME}
\label{appndx:analytic_1}

\newcommand{\caseoo}{Model case 1.1: Efficiency}
\addcontentsline{toc}{subsection}{\protect\numberline{}{\caseoo}}
\begin{description}
\item[\caseoo] \

We explore the dependence on characteristic scales using
a highly idealized model that satisfies the spatial and
temporal phase condition given by
\eq{eq:ft}, \eq{eq:fiip},  \eq{eq:muf_mu} and \eq{eq:ptype}.

\begin{enumerate}
\item Single mode $i$:
We consider a model:
\bea\label{eq:muCa_model_c}
\muCa\sbi(s,t)&=&\sigma\spC\sbi
\cos\left(\frac{\pi}{\SCi}s\right)
\cos\left(\frac{\pi}{\Tui}t\right)~,
\eea
where $\sigma\spC\sbi$ has a unit of (velocity)$^{2}$
corresponding to $\sigma\sbi |\vum(\vxCm(s))||\vu\sbi(\vxCm(s))|$
as in \eq{eq:must_s}.
A straightforward integration of \eq{eq:etstm_m} using
\eq{eq:muCa_model_c} gives an analytical form 
for $\aC\sbi(s,t)=\aC\sbi(s,t;s\spdgr:s,t\spdgr:t)$:
\bse\label{eq:muCa}
\bea\label{eq:muCa_m}
\aC\sbi(s,t)&=&
b\spC\sbi(t;s\spdgr,t\spdgr)+d\spC\sbi(t;s\spdgr,t\spdgr)
-[b\spC\sbi(t\spdgr;s\spdgr,t\spdgr)+d\spC\sbi(t\spdgr;s\spdgr,t\spdgr)]~,
\eea
where $s\spdgr=s-t+t\spdgr$, and
\bea\label{eq:muCa_b}
b\spC\sbi(t;s\spdgr,t\spdgr)&=&\sigma\spC\sbi\left\{ \begin{array}{ll}
t~\cos\{\frac{\pi}{\SCi}(s\spdgr-t\spdgr)\}~,
   & \mbox{for $\GCi=1$}~, \\
\frac{\SCi}{2\pi (1-\GCi)}
 ~\sin \{\frac{\pi}{\SCi}\left(1-\GCi\right)t
   + \frac{\pi}{\SCi}(s\spdgr-t\spdgr)\}~,
   & \mbox{for $\GCi\neq 1$}~: \end{array} \right.  \\
d\spC\sbi(t;s\spdgr,t\spdgr)&=& \begin{array}{l}
\sigma\spC\sbi\frac{\SCi}{2\pi (1+\GCi)}
 ~\sin \{\frac{\pi}{\SCi}\left(1+\GCi\right)t
   + \frac{\pi}{\SCi}(s\spdgr-t\spdgr)\}~, \end{array}
\label{eq:muCa_d}
\eea
\ese
are respectively the low- and high-frequency responses of TIME to mode $i$.
The last two terms of the right-hand side in \eq{eq:muCa_m}
are initial condition to satisfy $\aC(s,t\spdgr)=0$.

\item A dynamically coherent pair $[i,i+1]$:
Using $\SCi=S\spC\sbip$, instantaneous flux can be written in a form:
\bea\label{eq:muCa_model_pair}
\muCa\sbiip(s,t)&=&\left\{\begin{array}{rl}
 \sigma\spC\sbi\cos\left\{\frac{\pi}{\SCi}\left(s-\GCi t\right)\right\}~,
   & \mbox{for phase type (i)}~, \\
 \sigma\spC\sbi\cos\left\{\frac{\pi}{\SCi}\left(s+\GCi t\right)\right\}~,
   & \mbox{for phase type (ii)}~.
 \end{array} \right.
\eea
The corresponding TIME function is:
\bea\label{eq:muCa_m_pair}
\aC\sbiip(s,t)&=&\left\{\begin{array}{rl}
  2 b\spC\sbi(t;s\spdgr,t\spdgr)~-2b\spC\sbi(t\spdgr;s\spdgr,t\spdgr)~,
  & \mbox{for phase type (i)}~, \\
  2 d\spC\sbi(t;s\spdgr,t\spdgr)~-2d\spC\sbi(t\spdgr;s\spdgr,t\spdgr)~,
  & \mbox{for phase type (ii)}~.
 \end{array} \right.
\eea
Hence a \pair\ $[i,i+1]$ has either 
the low- or  high-frequency response to 
mode $i$  as in \eq{eq:muCa_b}, depending on the phase type.
\end{enumerate}
\end{description}

\newcommand{\caseot}{Model case 1.2: Geometry and evolution of
pseudo-lobes}
\addcontentsline{toc}{subsection}{\protect\numberline{}{\caseot}}
\begin{description}
\item[\caseot] \
Using the highly idealized model defined in Model case 1.1,
we consider the spatial coherency and temporal evolution of 
$\aC\sbiip(s,t)$.
\begin{enumerate}
\item Phase type (i) for low-frequency response:
Solving for $\aC\sbiip(s,t)=0$ gives the pseudo-PIP sequence:
\bse\label{eq:pyi}
\bea\label{eq:pyi_pip}
s\spC\sbmd{[i,i+1],j}(t)&=&
(j+\frac{1}{2})\SCi+\frac{1}{2}(1+\GCi)t-\frac{1}{2}(1-\GCi)t\spdgr~,
\eea
which is easily verified by a straightforward substitution.
Using \eq{eq:pA}, the signed area 
of pseudo-lobe  ${\cal L}\spC\sbijp(t)$ is given by:
\bea\label{eq:pyi_A}
A\spC\sbijp(t)&=&(-1)^{j} \sigma\spC\sbi
\left\{ \begin{array}{ll}
  \frac{2(t-t\spdgr)\SCi}{\pi}~,
    &\mbox{for $\GCi=1$}~, \\
 \frac{4(\SCi)^{2}}{\pi^{2} (1-\GCi)}
 \sin \{\frac{\pi(1-\GCi)}{2\SCi}(t-t\spdgr)\}~,
    &\mbox{for $\GCi\neq1$}~. \end{array} \right. 
\eea
\ese

\item Phase type (ii) for high-frequency response:
Similar to phase type (i) but using $d\spC(t;s\spdgr,t\spdgr)$ instead of 
$b\spC(t;s\spdgr,t\spdgr)$,  we obtain the following:
\bse\label{eq:pyii}
\bea\label{eq:pyii_pip}
s\spC_{[i,i+1],j}(t)&=&
(j+\frac{1}{2})\SCi+\frac{1}{2}(1-\GCi)t-\frac{1}{2}(1+\GCi)t\spdgr~; \\
A\spC\sbijp(t)&=&(-1)^{j} \sigma\spC\sbi
 \frac{4(\SCi)^{2}}{\pi^{2} (1+\GCi)}
 \sin \{\frac{\pi(1+\GCi)}{2\SCi}(t-t\spdgr)\}~; \label{eq:pyii_A}
\eea
\ese
\end{enumerate}
\end{description}

\subsection{Category 2: Definite TIME}
\label{appndx:analytic_2}

We explore the dependence on characteristic scales using
a highly idealized model that satisfies the 
temporal phase conditions given by
\eq{eq:ft} and \eq{eq:fiip}.
Spatial domain of integration for the $p$-th flux eddy is
\bea\label{eq:sasb_p}
(s_{a}:s_{b})&=&
\left\{\begin{array}{lclll}
(\sCidp-\frac{\SCidp}{2} & : &\sCidp-\frac{\SCidp}{2}) 
  & \mbox{for mode $i$} & \\
(\sCidp & : &\sCidp+{\SCidp}) 
  & \mbox{for mode $i+1$} & \mbox{phase type (i)} \\
(\sCidp-{\SCidp} & : &\sCidp) 
  & \mbox{for mode $i+1$} & \mbox{phase type (ii)} 
\end{array} \right.
\eea

\newcommand{\caseto}{Model case 2.1: One flux eddy}
\addcontentsline{toc}{subsection}{\protect\numberline{}{\caseto}}
\begin{description}
\item[\caseto] \

\begin{enumerate}
\item Single mode $i$:
We consider a highly idealized model for $p$-th flux eddy:
\bea\label{eq:mufi_mu}
\muCa\sbidp(s,t)&=&
\sigma\spC\sbidp~\cos\frac{\pi}{\SCidp}(s-\sCidp)\cos\frac{\pi}{\Tui}t~, 
 \qquad \mbox{for $|s-\sCidp|\leq \frac{\SCidp}{2}$}~.
\eea
It is straightforward to show that \eq{eq:etstm_m}
leads to an analytical form:
\bse\label{eq:mCi}
\bea\label{eq:mCi_st} 
\aC\sbidp(s,t)&=& M\spC\sbidp
  ~\cos\{\frac{\pi}{\Tui}(s-t-\sCidp)\}~,
\eea
where
\bea\label{eq:mCi_M} 
M\spC\sbidp&=&\left\{ \begin{array}{lcll}
\sigma\spC\sbidp\SCidp~
    &\mbox{for $\GCidp=1$}~, \\
\sigma\spC\sbidp\frac{4\SCidp}{\pi\{1-(\GCidp)^{2}\}}
  \cos\{\frac{\pi}{2}(1-\GCidp)\}
    &\mbox{for $\GCidp\neq1$}~. \end{array} \right.
\eea
\ese
Solving for $\aC\sbidp(s,t)=0$ in \eq{eq:mCi_st} gives the pseudo-PIP
sequence, which in turn leads to the signed area by \eq{eq:pA}:
\bse\label{eq:pLijp}
\bea\label{eq:pLijp_s}
s\spC\sbmd{i.p,j}(t)&=&(j+\frac{1}{2})\Tui+t+\sCidp~, \\
A\spC\sbmd{i.p,j}(t)&=&(-1)^{j}M\spC\sbidp\frac{2\Tui}{\pi}~.
\label{eq:pLihp_A} 
\eea
\ese

\item A \pair\ $[i,i+1]$:
A complementary model for mode $i+1$ to \eq{eq:mufi_mu} 
using $S\spC\sbmd{i+1,p}=\SCidp$,
$\Gamma\spC\sbipdp=\GCidp$ and $\sigma\spC\sbidp=\sigma\spC\sbipdp$ is:

\bea\label{eq:mufip_mu}
\muCa\sbipdp(s,t)&=&
\sigma\spC\sbidp
  ~\cos\frac{\pi}{\SCidp}(s-\sCidp\pm\frac{\SCidp}{2})\sin\frac{\pi}{\Tui}t~
 \qquad 
 \mbox{for $|s-\sCidp\pm\frac{\SCidp}{2}|\leq \frac{\SCidp}{2}$}~, 
\eea
where $+$ and $-$ correspond to phase types (i) and (ii),
respectively.
It is straightforward to show that \eq{eq:etstm_m} leads to 
\bea\label{eq:mCidp_st} 
\aC\sbipdp(s,t)&=&
  M\spC\sbidp~\cos\{\frac{\pi}{\Tui}(s-t-\sCidp
 + \theta\spC\sbiipdp)\}~,
\eea
where 
\bea
\theta\spC\sbiipdp&=&\frac{\SCidp}{2}(1\mp\GCidp)
\eea
is the phase shift from mode $i$ 
with $-$ and $+$ for phase types (i) and (ii), respectively.
As a \pair\ $[i,i+1]$, 
we obtain from \eq{eq:mCi_st} and \eq{eq:mCidp_st}:
\bse\label{eq:mCiip}
\bea
\aC\sbiipdp(s,t) 
&=&M\spC{\sbiipdp}
\cos\{\frac{\pi}{\Tui}(s-t-\sCidp+\frac{\theta\spC\sbiipdp}{2})\}~,
  \label{eq:mCiip_st} 
\eea
 where
\bea\label{eq:mCiip_M}
M\spC\sbiipdp = 2~M\spC\sbidp~
   \cos\{\frac{\pi}{4}\GCidp(1\mp\GCidp)\}~.
\eea
Solving for $\aC\sbiipdp(s,t)=0$ in \eq{eq:mCiip_st} gives
the pseudo-PIPs, which in turn lead to the signed area:
\bea\label{eq:pLiijp_s}
s\spC\sbmd{[i,i+1].p,j}(t)&=&
  (j+\frac{1}{2})\Tui+t+\sCidp-\frac{\theta\spC\sbiipdp}{2}~,\\
A\spC\sbmd{[i,i+1].p,j}(t)&=&(-1)^{j}M\spC\sbiipdp\frac{2\Tui}{\pi}~.
\label{eq:pLiihp_A} 
\eea
\ese
\end{enumerate}
\end{description}

\newcommand{\casett}{Model case 2.2: Entire $C$}
\addcontentsline{toc}{subsection}{\protect\numberline{}{\casett}}
\begin{description}
\item[\casett] \

\begin{enumerate}
\item Idealized model for $\muC\sbi(s)$.
By modeling $\muC\sbi(s)$ as $N\spC\sbi$ idealized flux eddies 
individually defined by \eq{eq:mufi_mu},
we obtain $\aC\sbi(s,t)$ in a form of:
\bse\label{eq:mC}
\bea\label{eq:mC_m}
\aC\sbi(s,t)
&=& M\spC\sbi \cos
\{\frac{\pi}{\Tui}(s-t-s\spC\sbi+\theta\spC\sbi)\}~
\eea
similar to \eq{eq:mCi_st} for $\aC\sbidp(s,t)$.
The amplitude and phase shift
\bea
M\spC\sbi &=&\sqrt{
  \{m\spC\sbmd{i,\sin}\}^{2}
 +\{m\spC\sbmd{i,\cos}\}^{2}}~: \label{eq:mC_M} \\
\theta\spC\sbi &=& \tan^{-1}
  \left(\frac{m\spC\sbmd{i,\sin}}{m\spC\sbmd{i,\cos}}\right)~.
\label{eq:mC_a}
\eea
include contribution from all individual flux eddies, where
\bea
m\spC\sbmd{i,\sin}&=&
 \sum_{p=1}^{N\spC\sbi}M\spC\sbidp~\sin\frac{\pi}{\Tui}\sCidp~,\\
m\spC\sbmd{i,\cos}&=&
\sum_{p=1}^{N\spC\sbi}M\spC\sbidp~\cos\frac{\pi}{\Tui}\sCidp~.
\label{eq:msfcos_cs}
\eea
\ese
Solving for $\aC\sbidp(s,t)=0$ in \eq{eq:mC_m} gives
the pseudo-PIPs, which in turn lead to the signed area:
\bse
\bea\label{eq:msA_s}
s\spC\sbmd{i,j}(t)&=&(j+\frac{1}{2})\Tui+t+\theta\spC\sbi~~,\\
A\spC\sbmd{i,j:j+1}(t)&=&(-1)^{j}\frac{2  M\spC\sbi\Tui}{\pi}~.
\label{eq:msA_A}
\eea
\ese
\item General form of $\muC\sbi(s)$.
Using $\muC\sbi(s)$ over a segment $[s_{a},s_{b}]$ of $C$,
it is fairly straightforward to show that $\aC\sbi(s,t)$ has the same
form as \eq{eq:mC} by setting $s\spC\sbmd{i,1}=0$ 
and replacing \eq{eq:msfcos_cs} with 
\bse\label{eq:msf}
\bea
m\spC\sbmd{i,\sin}&=&
 \int_{s_{a}}^{s_{b}}\muC\sbi(\tau)\sin\frac{\pi}{\Tui}\tau ~d\tau~,\\
m\spC\sbmd{i,\cos}&=&
 \int_{s_{a}}^{s_{b}}\muC\sbi(\tau)\cos\frac{\pi}{\Tui}\tau ~d\tau~.
\label{eq:msf_cs}
\eea
\ese
\end{enumerate}

\end{description}

\clearpage
\addcontentsline{toc}{section}{\protect\numberline{}{References}}
\bibliographystyle{natbib}

\clearpage

\renewcommand{\thesection}{}
\section{Tables}
\begin{table}[!ht]
\begin{center}
\scalebox{0.9}{
\begin{tabular}{|rr||c|rr||c||c|c|c||c|c|c||c|c|} \hline
\multicolumn{2}{|c||}{} 
  & \multicolumn{3}{c||}{}
  & \multicolumn{9}{c|}{characteristic scale} \\ \cline{6-14}
\multicolumn{2}{|c||}{mode} 
  & \multicolumn{3}{c||}{wave}
  & temporal & \multicolumn{8}{c|}{spatial} \\ \cline{7-14}
\multicolumn{2}{|c||}{ } 
  & \multicolumn{3}{c||}{ }
  & & \multicolumn{3}{c||}{dynamic}
    & \multicolumn{3}{c||}{global flux} 
    & \multicolumn{2}{c|}{flux alignment}  \\ \hline
$i$ & $i+1$
  & $n$ & $(\knb,$ & $\lnb)$ 
  & $T\spu\sbi$
  & $N\spu\sbi$ & $L\spu\sbi$ & $b\spu\sbi$ 
  & $N\spp\sbi$ & $L\spp\sbi$ & $b\spp\sbi$ 
  & $S\spp\sbi$ & $\Gamma\spp\sbi$ \\ \hline \hline
1, & 2
  & 1 & (1, & 2 )
  & 1.66667
  & 1 & 1  & 0.6 & 2 & 0.5 & 0.3 & 0.3151 & 0.189 \\ \hline
3,& 4 
  & 2 & (2, & 1 )
  & 0.83333
  & 2 & 0.5 & 0.6 & 2 & 0.5 & 0.6 & 0.3151 & 0.3781 \\ \hline
5,& 6 
  & 3 & (2, & 2 )
  & 0.66667
  & 2 & 0.5 & 0.75 & 3 & 0.3333 & 0.5
                   & 0.2101 & 0.3151 \\ \cline{10-14}
  & 
  &   &   & 
  & 
  &   &     &      & 1$\spstr$ & 1$\spstr$ & 1.5$\spstr$ 
                   & 0.6303 $\spstr$ & 0.9454$\spstr$ \\ \hline
7,& 8 
  & 4 & (2, & 3 )
  & 0.59090
  & 2 & 0.5 & 0.84 & 4 & 0.25 & 0.42 & 0.15 & 0.26 \\ \hline
9,& 10 
  & 5 & (3, & 2 )
  & 0.39393
  & 3 & 0.3333 & 0.84 & 4 & 0.25 & 0.63 & 0.15 & 0.39 \\ \hline
\end{tabular}
}
\caption{Characteristic scales of the atmospheric model.
The variance is $\sigma\sbi=0.1$ is for all ten modes
(five \RTs).
For modes 5 and 6 of the third \RT,
the case with $N\spp\sbwv{3}=1\spstr$ is also listed with 
the superscript.
}
\label{tbl:rt}
\end{center}
\end{table}

\begin{table}[!ht]
\begin{center}
\begin{tabular}{|rr||c||cc||c|c||c|c|} \hline
\multicolumn{2}{|c||}{ } & 
  & \multicolumn{2}{|c||}{}
  & \multicolumn{4}{c|}{response to variability} \\ \cline{6-9}
\multicolumn{2}{|c||}{mode} & wave 
  & \multicolumn{2}{|c||}{spatial bias}
  & \multicolumn{2}{c||}{low-frequency}
  & \multicolumn{2}{c|}{high-frequency} \\ \hline
$i$ & $i+1$ & $n$
  & $F\spC\sbi$ & $(i)$ 
  & $S\spC/(1-\Gamma\spC\sbi)$ & $(1+\Gamma\spC\sbi)/2$ 
  & $S\spC/(1+\Gamma\spC\sbi)$ & $(1-\Gamma\spC\sbi)/2$ \\ \hline\hline
1, & 2 & 1
  &  -14.4841 & $(2)$
  & 0.3888 &   0.5945 &  0.26503 &  0.40546 \\ \hline
3, & 4 & 2
  &  -4.3300 & $(3)$
  & 0.5068  & 0.6891  &  0.2287 &  0.3109 \\ \hline
5, & 6 & 3
  &    & 
  & 0.3068 & 0.6576 & 0.1598 &  0.3424\\ \cline{6-9}
   &   &  
  &  & 
  &  11.5456 $\spstr$ & 0.9727 $\spstr$ & 0.32398 $\spstr$ &
	0.02729$\spstr$ \\ \hline
7, & 8 & 4
  &  0.6745 & $(7)$
  & 0.2149 & 0.6333 & 0.1244 &  0.3667\\ \hline
9, & 10 & 5
  &  4.6642 & $(10)$
  & 0.2626 & 0.7000 & 0.1126  & 0.3000 \\ \hline
\end{tabular}
\caption{The spatial bias of the instantaneous flux and 
properties concerning 
low- and high-frequency responses of finite-time TIME
along $\Ca$ for the ten modes (five waves) as in \tbl{tbl:rt}.
$F\spC\sbi$ is listed only for those with non-zero value.
For modes 5 and 6 of the third \RT,
the results with $N\spp\sbwv{3}=1\spstr$ are also listed with 
the superscript.
}
\label{tbl:rt_Ca}
\end{center}
\end{table}
\clearpage

\section{Figures}

\begin{figure}[htb!]
\begin{center}
\includegraphics[width=18cm]{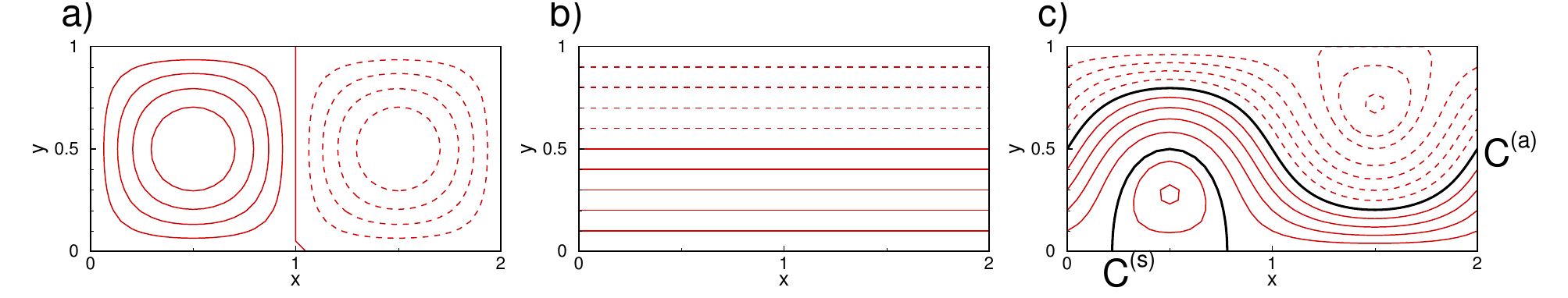}
\end{center}
\caption{Reference streamfunction fields with solid and broken 
lines for positive and negative contour values:
a) principal Rossby wave; b) eastward jet; and
c) total for a supercritical case with $\azb=0.5$.
Two thick lines in c are the kinematically defined boundary curves used in this study:
a periodic orbit $\Ca$ corresponding to the jet axis with period
$T^{(a)}=1.260549$;  
a separatrix $\Cs$  dividing the southern anti-cyclonic
recirculating cell from the eastward jet.
The trough and ridges of the reference flow are at $x=1/2$ and
$3/2$, respectively.
}
\label{fg:rt_rf}
\end{figure}

\begin{figure}[htb!]
\begin{center}
\includegraphics[height=13cm]{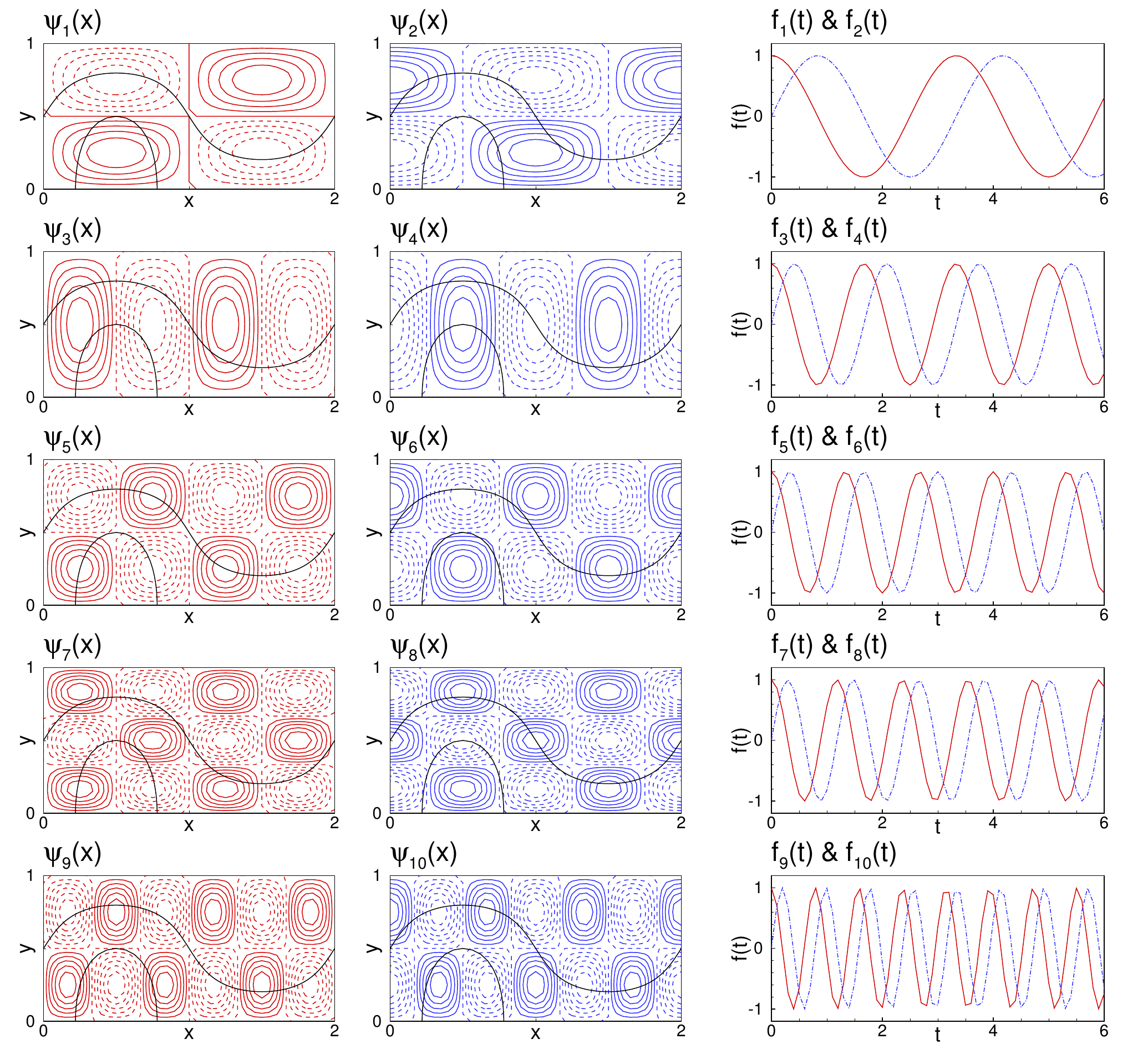}
\end{center}
\caption{Spatial and temporal components of the dynamical modes
corresponding to the five \RT s  (see \tbl{tbl:rt}).
The left and center panels are $\psi\sbtnn(\vx)$ and $\psi\sbtn(\vx)$,
respectively, where solid and dashed lines represent positive and
negative contour values with contour interval 0.2.
The two boundary curves $\Ca$ and $\Cs$ are also shown by the
solid curves.
The right panels are $f\sbtnn(t)$ and $f\sbtn(t)$
in solid and dashed lines, respectively.
}
\label{fg:rt_psia}
\end{figure}

\begin{figure}[!ht]
\begin{center}
\includegraphics[height=13cm]{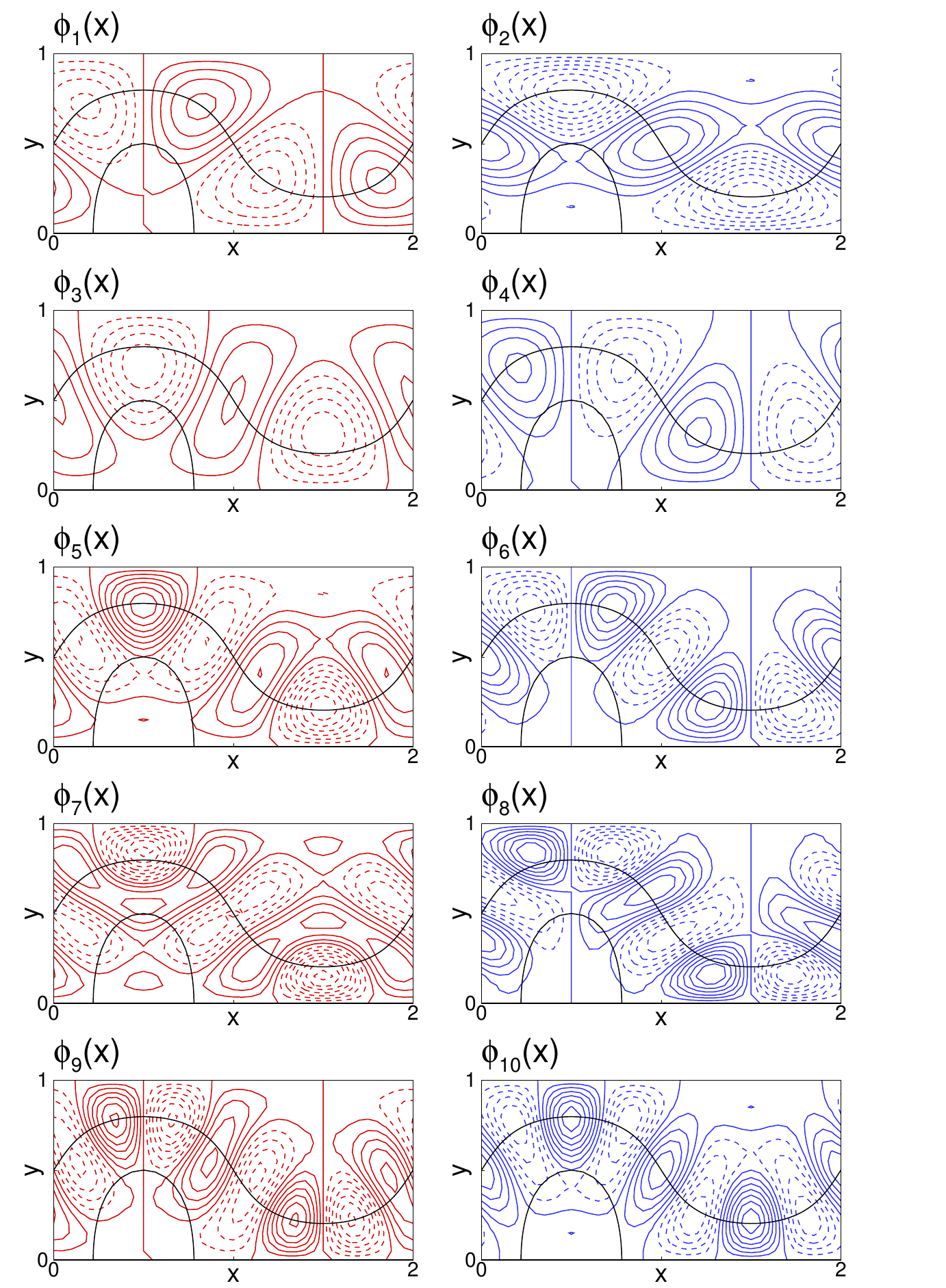}
\end{center}
\caption{Spatial components $\phi\sbi(\vx)$  of the flux mode
corresponding to \fg{fg:rt_psia} ,
where solid and dashed lines represent positive and
negative contour values with contour interval 1.
}
\label{fg:rt_phia_x}
\end{figure}

\begin{figure}[!ht]
\begin{center}
\includegraphics[width=14cm]{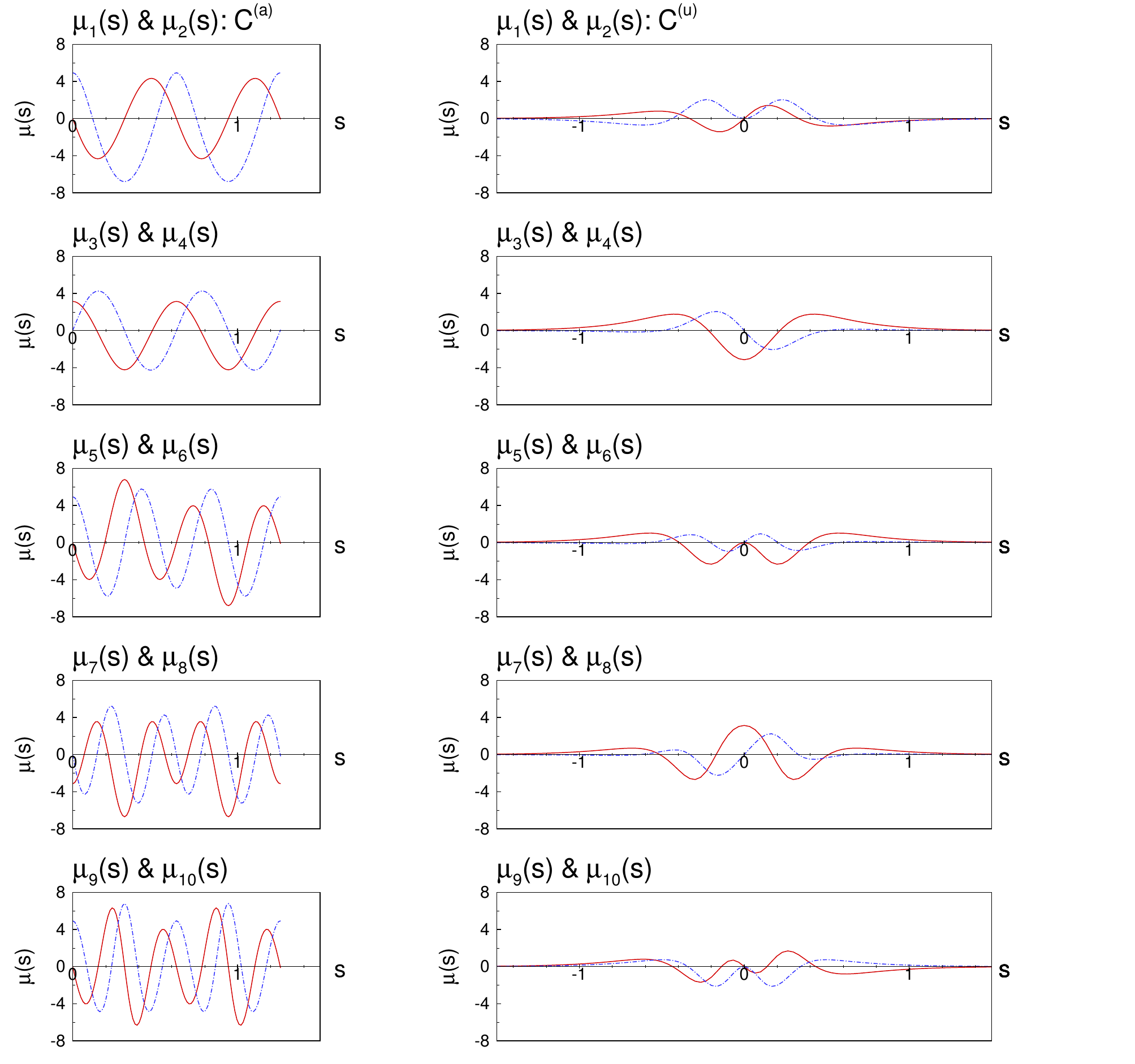}
\end{center}
\caption{Spatial component $\muC\sbi(s)$ of flux mode along 
$\Ca$ (left panels) and $\Cs$ (right panels) corresponding to  
\fg{fg:rt_phia_x}, where solid and dashed lines are for
$\muC\sbtnn(s)$ and $\muC\sbtn(s)$, respectively.
In the left panels, $s=0.3157$ and $0.9454$ correspond to 
the two turning points of $\Ca$ at $x=1/2$ and $3/2$
over the trough and ridge.
In the right panels, $s=0$ corresponds to the trough along 
$\Cs$. 
All flux eddies of $\mu\sps\sbi(s)$ concentrate
over the segment $s\in[-1,1]$ where 
${\bar \vx}\sps(-1)$ and ${\bar \vx}\sps(1)$ 
are near the upstream and downstream DHTs 
${\bar \vx}\sps(-\infty)$ and ${\bar \vx}\sps(\infty)$ 
in the $\vx$ space, respectively.
}
\label{fg:rt_phia_s}
\end{figure}


\begin{figure}[!ht]
\begin{center}
\includegraphics[height=18cm]{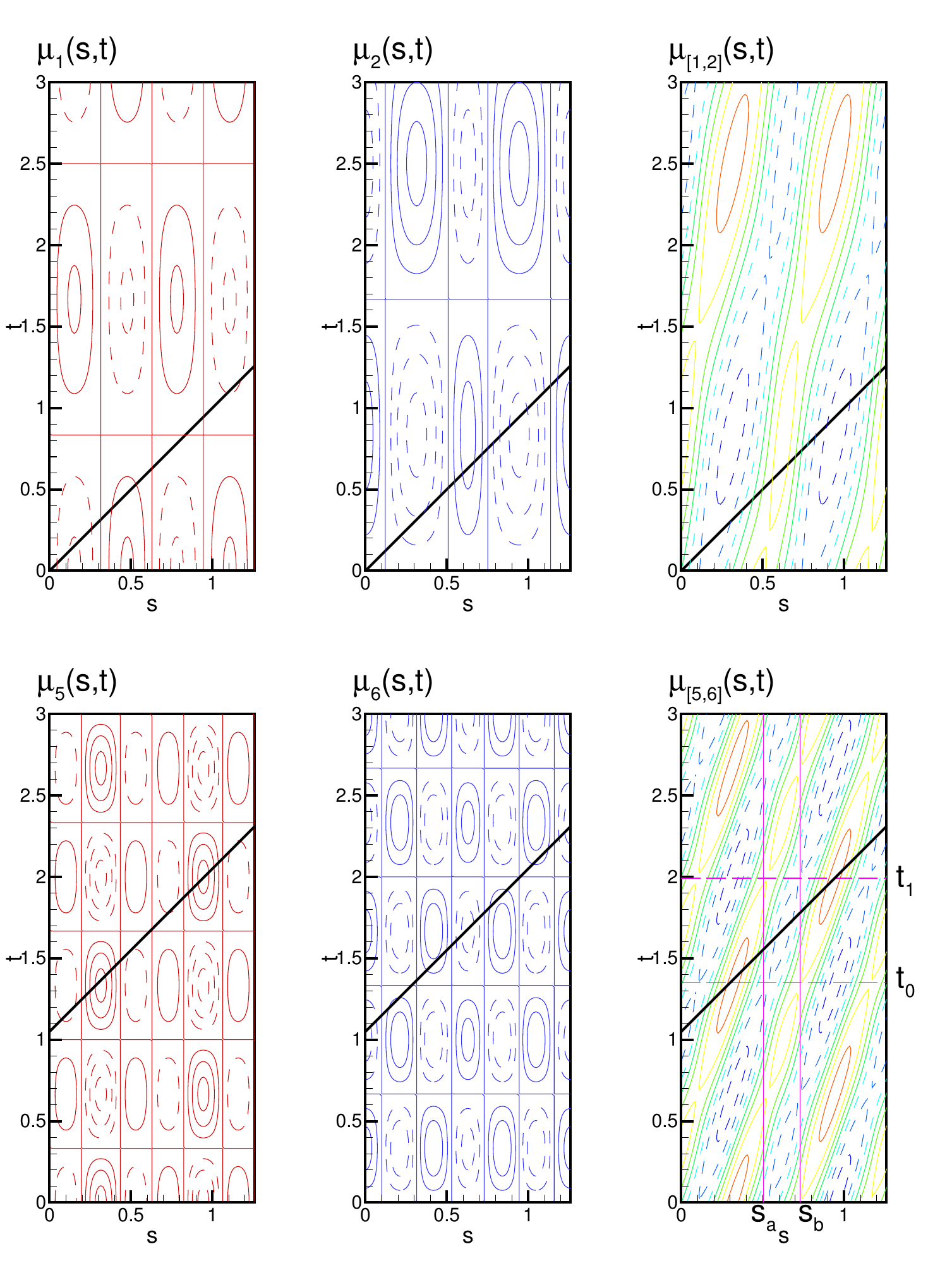}
\end{center}
\caption{Instantaneous flux $\muaa(s,t)$ along $\Ca$ 
corresponding to Figures \ref{fg:rt_phia_x} and \ref{fg:rt_phia_s}
where the solid and dashed lines are for positive and negative contour 
values with the contour interval 0.2;
top panels are
$\muaa\sbmd{1}(s,t)$, $\muaa\sbmd{2}(s,t)$, and
$\muaa\sbmd{[1,2]}(s,t)$,
and bottom panels are
$\muaa\sbmd{5}(s,t)$, $\muaa\sbmd{6}(s,t)$, and
$\muaa\sbmd{[5,6]}(s,t)$.
An example for a diagonal line of integration $(s-t+\tau,\tau)$ going
through  $(0.625,1.675)$ is shown in each panel;
an example for domain of integration 
$(s_{a}:s_{b},t_{0},t_{1})=(0.51:0.74,1.35:2)$
is shown in $\muaa\sbmd{[5,6]}(s,t)$.
}
\label{fg:rt_must_axis}
\end{figure}

\begin{figure}[!ht]
\begin{center}
\includegraphics[height=10cm]{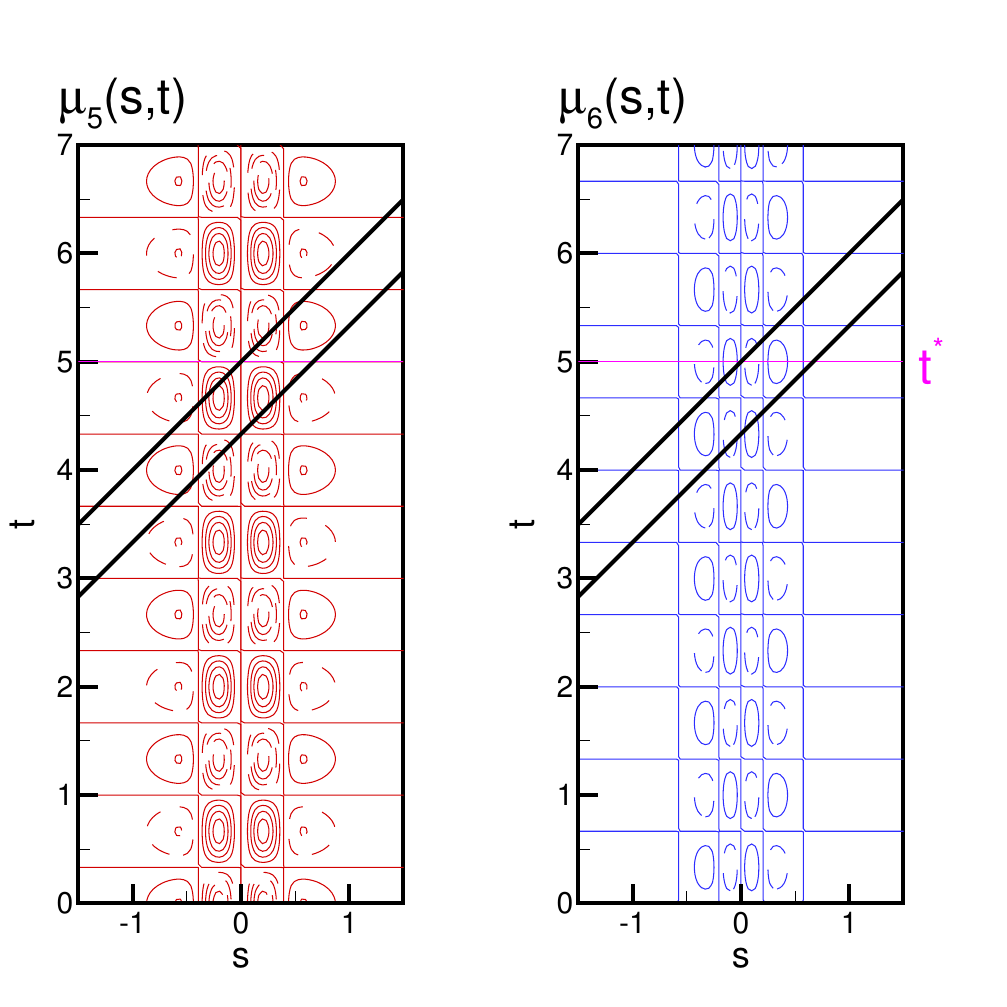}
\end{center}
\caption{Instantaneous flux
$\musa\sbmd{5}(s,t)$ and $\musa\sbmd{6}(s,t)$
for modes 5 and 6 along $\Cs$ corresponding to Instantaneous flux
$\musa\sbmd{5}(s,t)$ and $\musa\sbmd{6}(s,t)$
for modes 5 and 6 along $\Cs$ corresponding to 
Figures \ref{fg:rt_phia_x} and \ref{fg:rt_phia_s}
where the solid and dashed lines are for positive and negative contour 
values with the contour interval 0.05;
Two example reference trajectories $(s-t+\tau,\tau)$ are
separated by $T\spu\sbmd{5}$ in $s$ and $t$.
An example evaluation time $t^{*}=5$ is also shown in the right panel.
}
\label{fg:rt_must_cellu}
\end{figure}

\begin{figure}[!ht]
\begin{center}
\includegraphics[width=14cm]{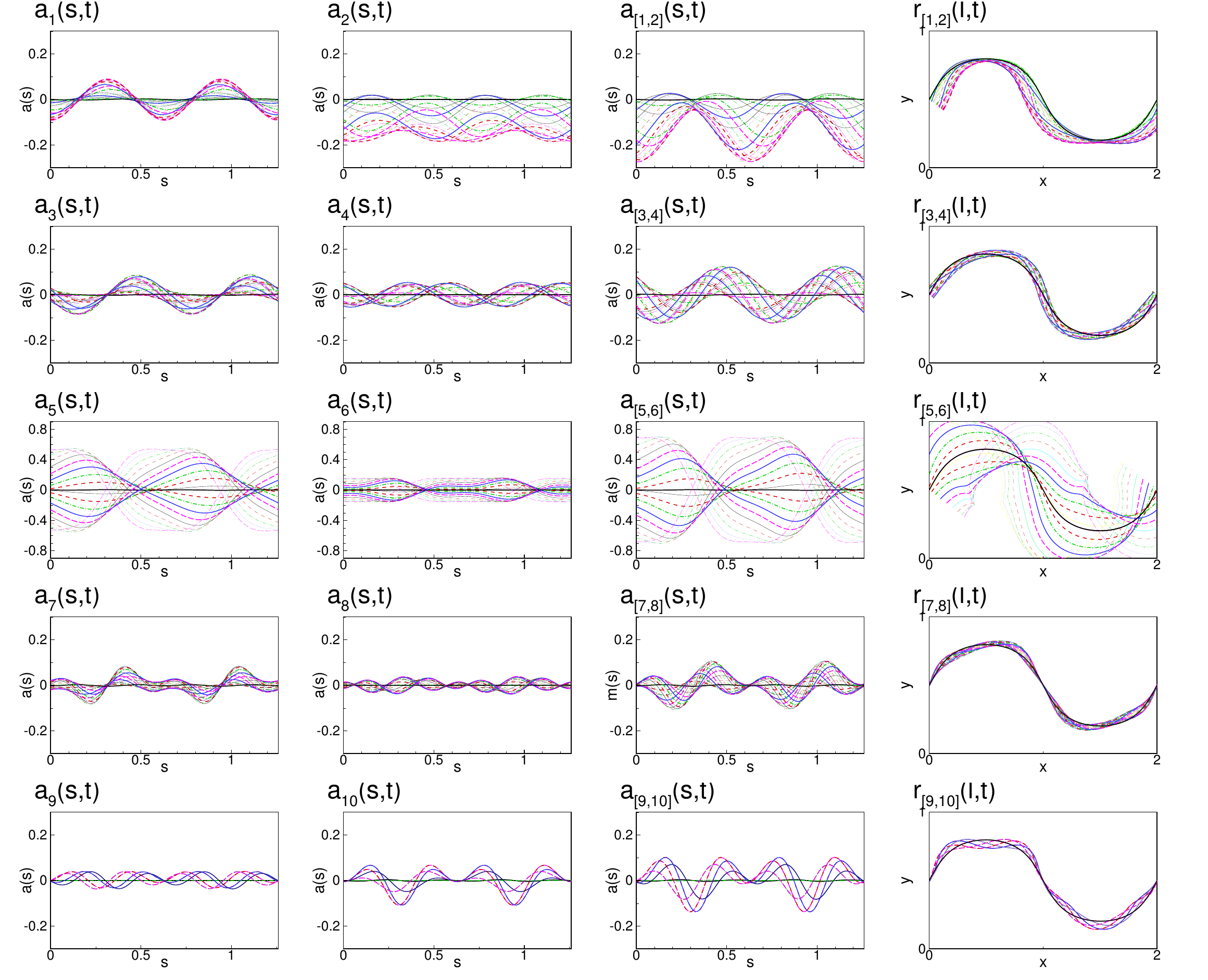}
\end{center}
\caption{Envelopes of $\aa\sbtnn(s,t)$, $\aa\sbtn(s,t)$ and
$\aa\sbmd{[2n-1,2n]}(s,t)$  in the $(s,a)$ space
(panels in three left columns, respectively) and 
$r\spa\sbmd{[2n-1,2n]}(l,t)$ in the $\vx$ space
(panels in right column) along $\Ca$ 
where 20 envelopes are taken at every one half $T\spa$.
The scales of $\aa$ for $\aa\sbmd{5}(s,t)$, $\aa\sbmd{6}(s,t)$, 
and $\aa\sbmd{[5,6]}(s,t)$ differ from others
due to large amplitude, while $r\spa\sbmd{[5,6]}(s,t)$ is unscaled 
for comparison with pseudo-lobes of other pairs in $\vx$.
}
\label{fg:rt_ms_axis_range}
\end{figure}

\begin{figure}[!ht]
\begin{center}
\includegraphics[width=14cm]{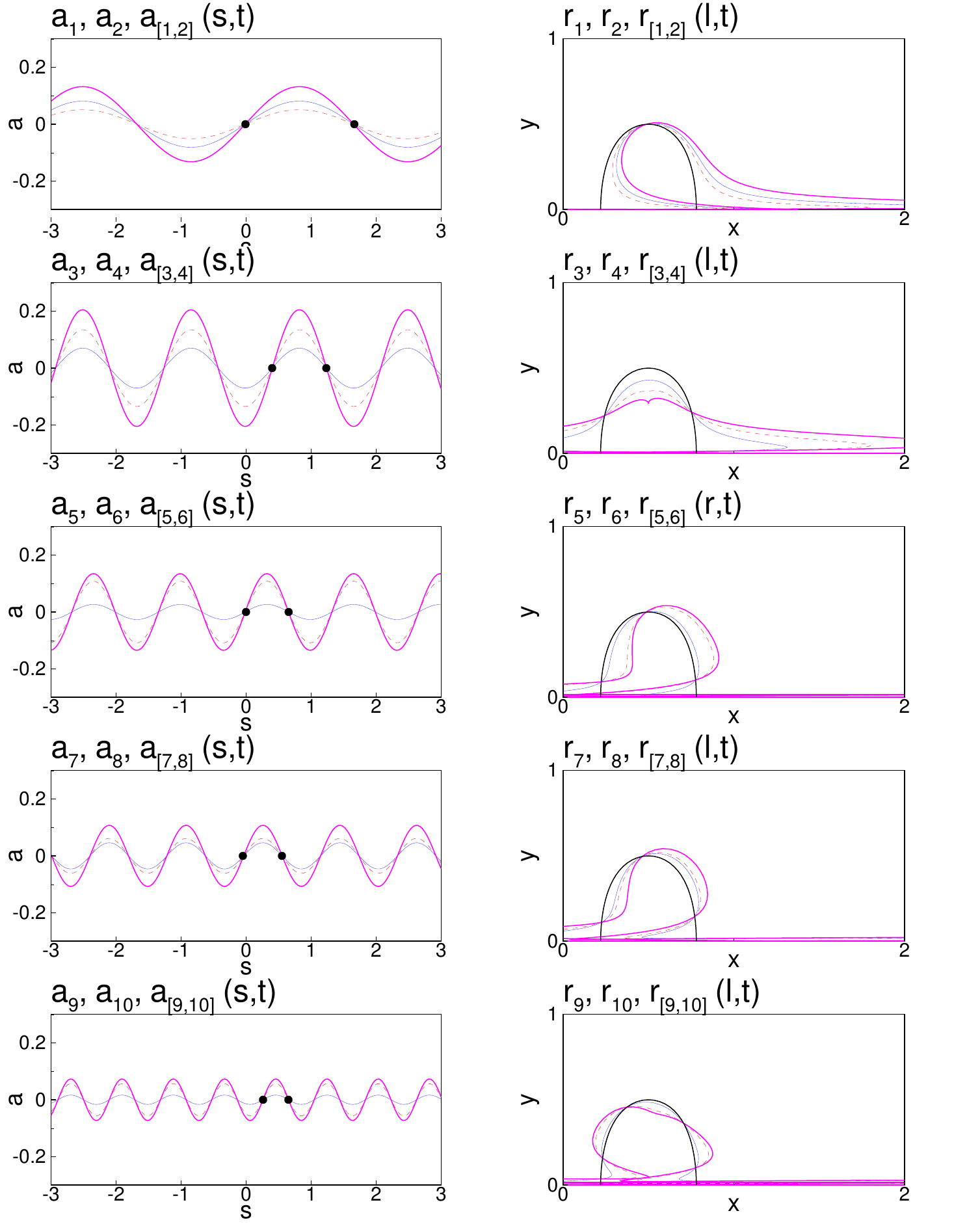}
\end{center}
\caption{$\as\sbtnn(s,t)$, $\as\sbtn(s,t)$ and
$\as\sbmd{[2n-1,2n]}(s,t)$ in the $(s,a)$ space (left panels) and 
$r\sps\sbtnn(l,t)$, $r\sps\sbtn(l,t)$ and 
$r\sps\sbmd{[2n-1,2n]}(l,t)$ in the $\vx$ space
(right panels) along $\Cs$ at $t\spstr=5$;
solid and dashed lines are for even and odd modes, respectively,
and thick line is for the pair.
The circles are the upstream pseudo-PIP at $s\spstr$ 
downstream pseudo-PIP at $s\spstr+T\spu\sbmd{2n-1}$
for the positive pseudo-lobe ${\cal L}\spstr$.
The pseudo-PIPs correspond to the diagonal lines
in \fg{fg:tr_must_cellu}.
}
\label{fg:rt_ms_cellu_t5}
\end{figure}

\begin{figure}[!ht]
\begin{center}
\includegraphics[height=18cm]{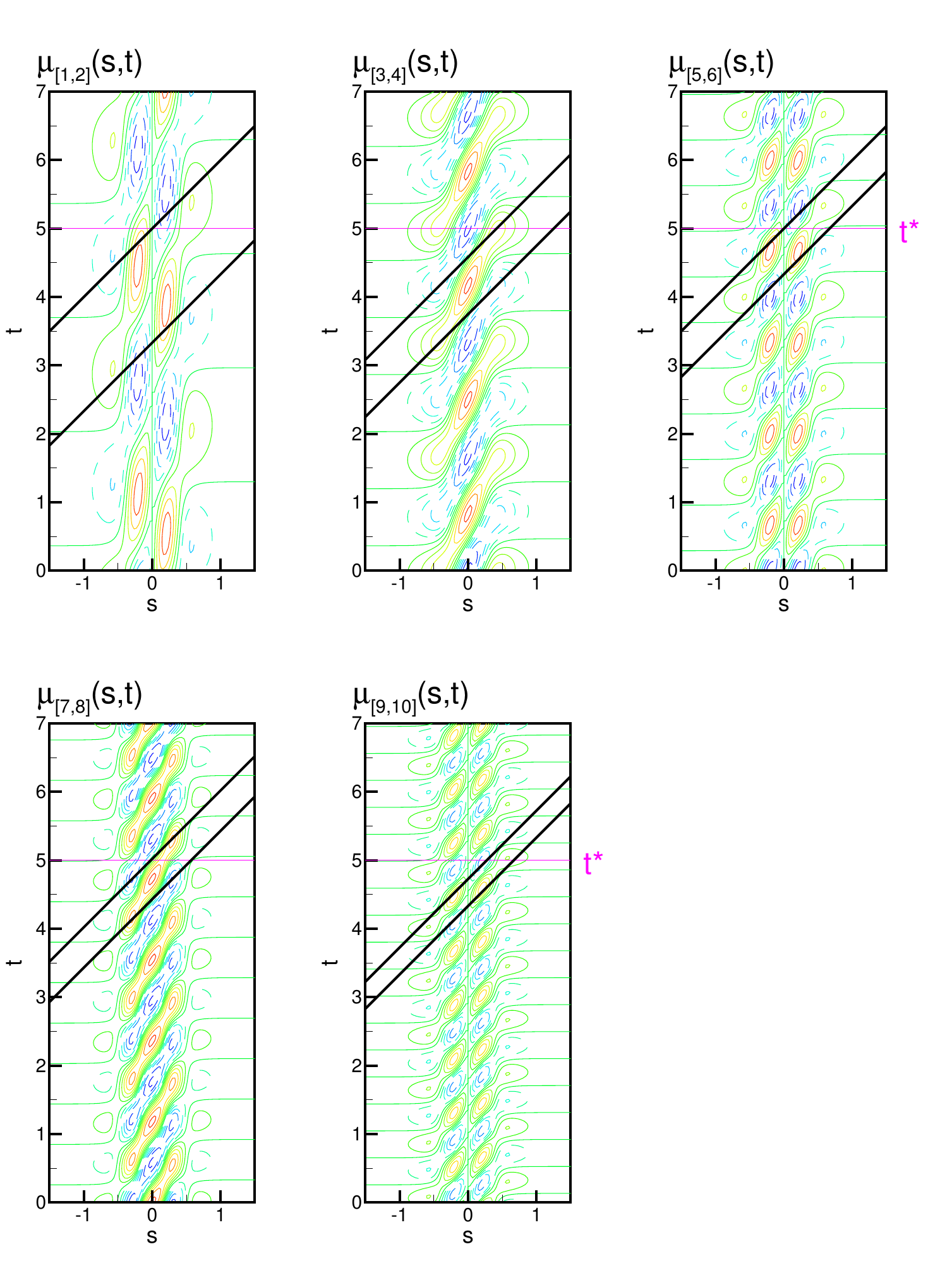}
\end{center}
\caption{Instantaneous flux
$\musa\sbmd{[2n-1,2n]}(s,t)$ for the five \RT s along $\Cs$.
The two diagonal lines correspond to the two pseudo-PIPs
of the pseudo-lobe ${\cal L}\spstr$,
as indicated by the circles in \fg{fg:rt_ms_cellu_t5};
$t\spstr$ is the evaluation time.
}
\label{fg:tr_must_cellu}
\end{figure}

\begin{figure}[!ht]
\begin{center}
\includegraphics[height=10cm]{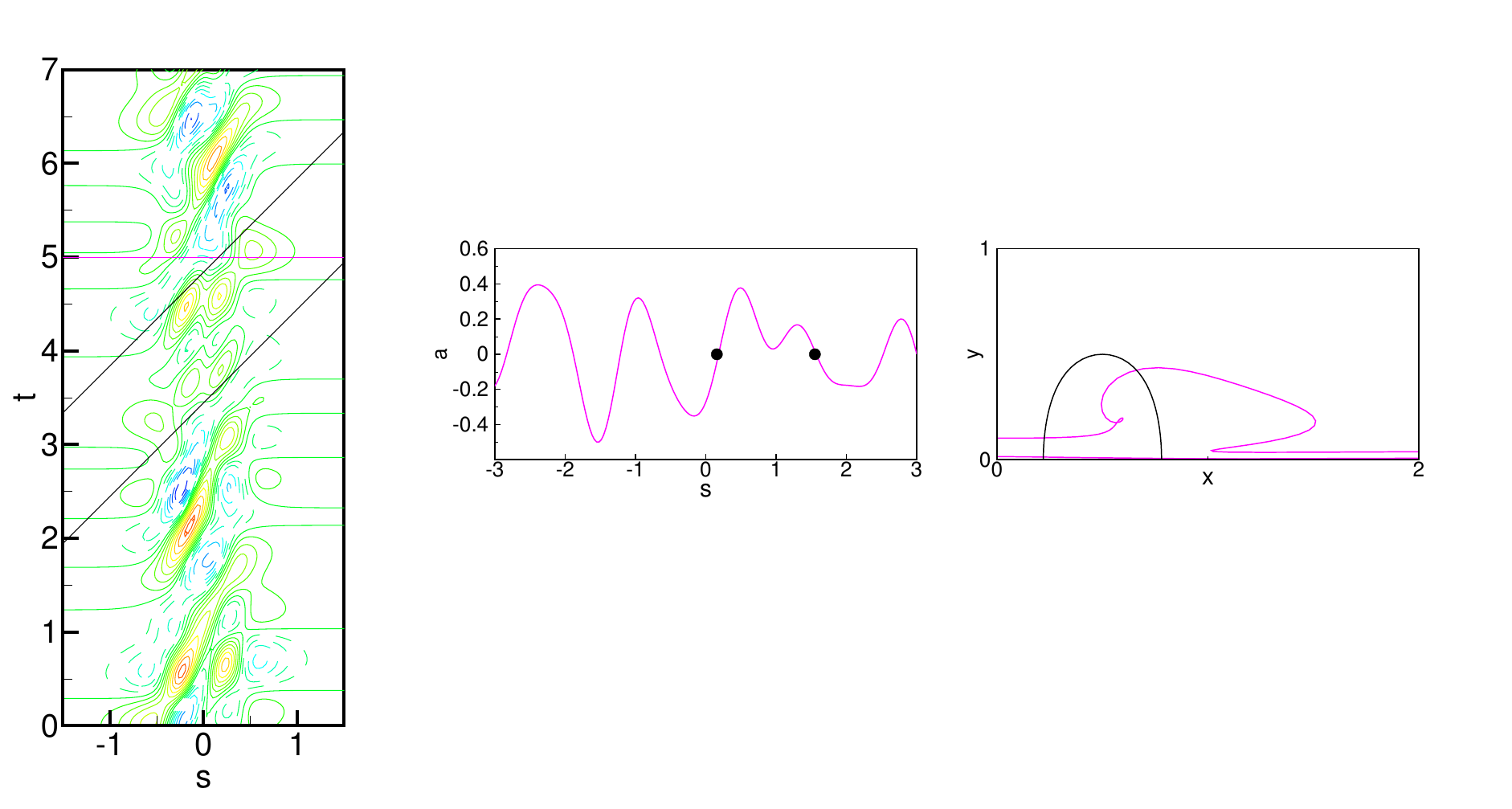}
\end{center}
\caption{Same as Figures \ref{fg:rt_ms_cellu_t5}
and \ref{fg:tr_must_cellu}, but 
for all five Rossby waves together.}
\label{fg:tr_must_cellu_t5}
\end{figure}

\begin{figure}[!ht]
\begin{center}
\includegraphics[height=10cm]{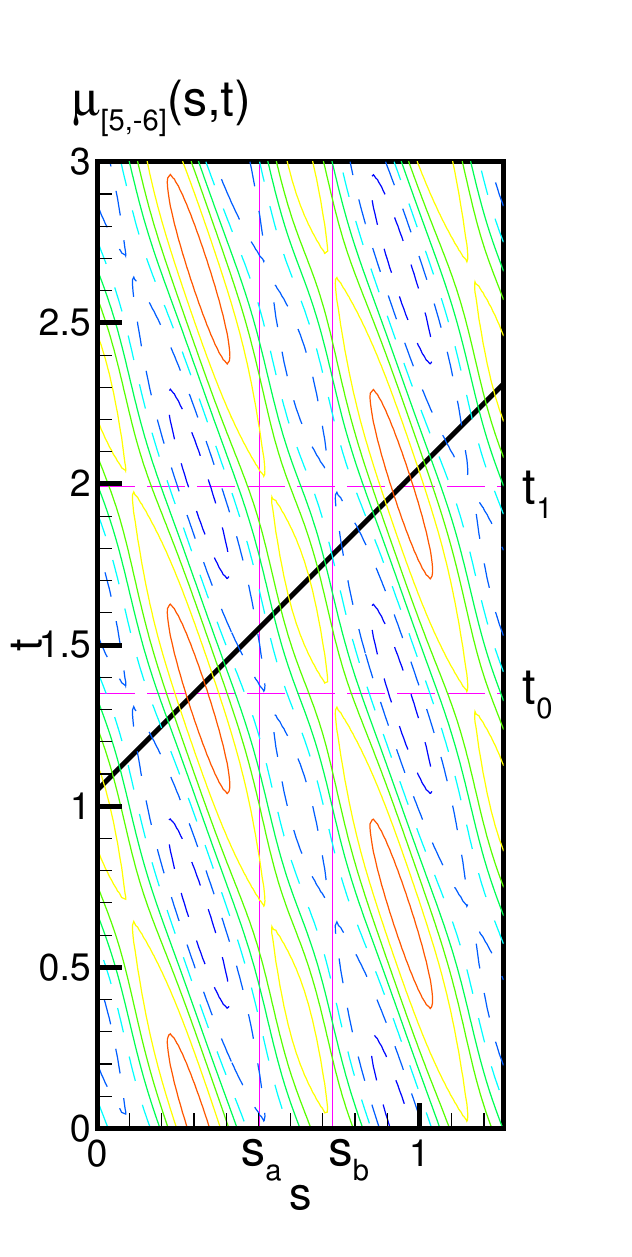}
\end{center}
\caption{Instantaneous flux $\muaa\sbmd{[5,-6]}(s,t)$ 
along $\Ca$ for phase type (ii), corresponding to upstream
propagation of the flux eddies as a counterpart of
$\muaa\sbmd{[5,6]}(s,t)$ in \fg{fg:rt_must_axis}.
}
\label{fg:rt22rv_axis}
\end{figure}

\begin{figure}[!ht]
\begin{center}
\includegraphics[height=13cm]{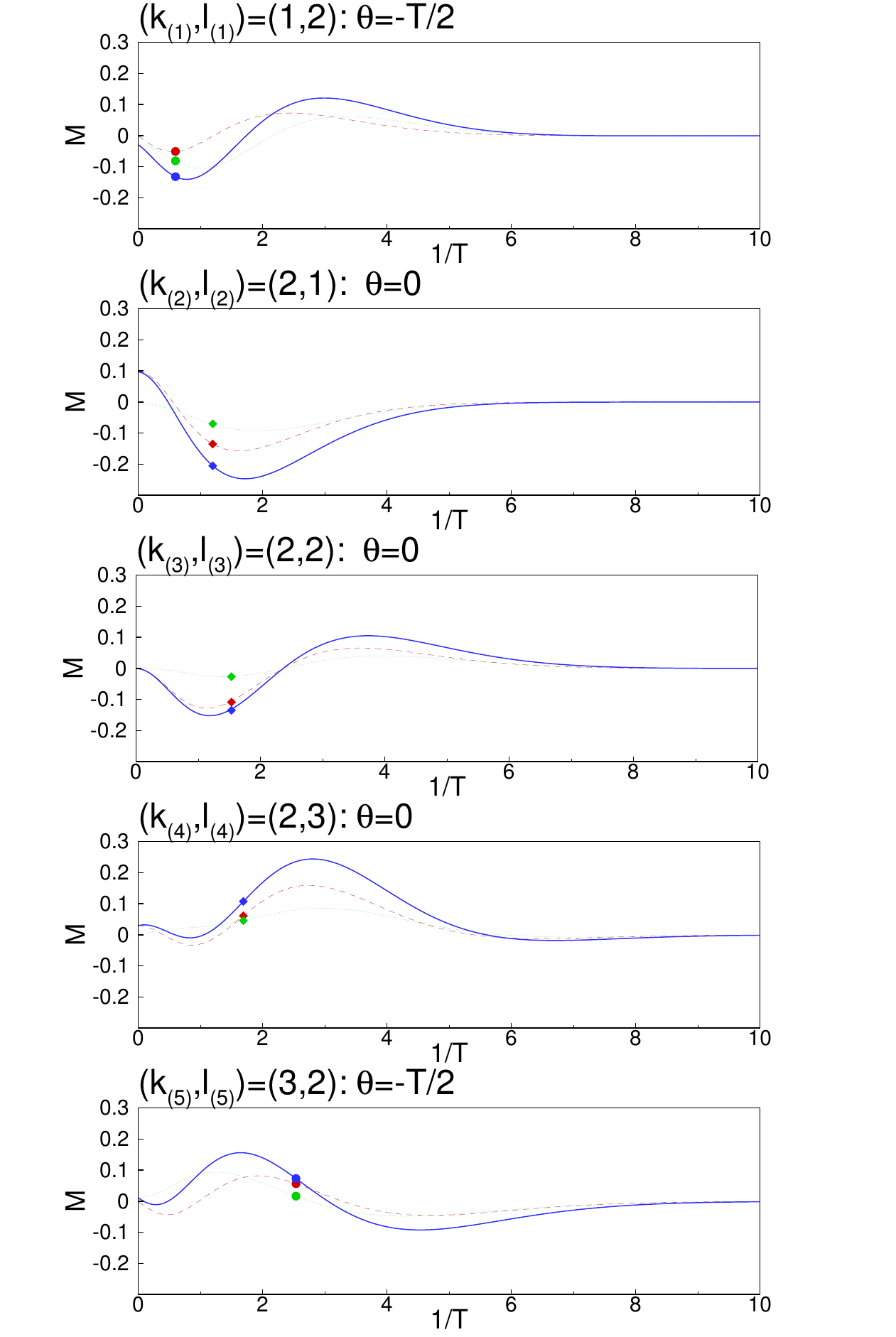}
\end{center}
\caption{$M\sps\sbd$ as function of 
$1/T$
using the spatial components $\mu\sps\sbtnn(s)$ and 
$\mu\sps\sbtn(s)$ along $\Cs$ but $f(t)=\cos\pi t/T$ 
instead of $f\sbtnn(t)$ and $f\sbtn(t)$, respectively.
The symbols corresponding to $T=\Tui$ for \model.}
\label{fg:rt_Mcs}
\end{figure}

\end{document}